\documentclass[10pt,aps,prd,fleqn,superscriptaddress,notitlepage,nofootinbib,preprintnumbers,nobalancelastpage]{revtex4-1}
\usepackage{amsmath,amssymb,graphicx,xspace,subfigure,siunitx,commath}
\usepackage{todonotes}
\newcommand{\Sherpa}{S\protect\scalebox{0.8}{HERPA}\xspace}
\newcommand{\Pythia}{P\protect\scalebox{0.8}{YTHIA}\xspace}
\newcommand{\Djangoh}{D\protect\scalebox{0.8}{JANGOH}\xspace}
\newcommand{\RapGap}{R\protect\scalebox{0.8}{AP}G\protect\scalebox{0.8}{AP}\xspace}
\newcommand{\Ariadne}{A\protect\scalebox{0.8}{RIADNE}\xspace}
\newcommand{\BlackHat}{B\protect\scalebox{0.8}{LACK}H\protect\scalebox{0.8}{AT}\xspace}
\newcommand{\Apfel}{A\protect\scalebox{0.8}{PFEL}\xspace}
\newcommand{\Rivet}{R\protect\scalebox{0.8}{IVET}\xspace}
\newcommand{\Amegic}{A\protect\scalebox{0.8}{MEGIC}\xspace}
\newcommand{\Dire}{D\protect\scalebox{0.8}{IRE}\xspace}
\newcommand{\unlops}{U\protect\scalebox{0.8}{NLOPS}\xspace}
\newcommand{\mcatnlo}{S-\protect\scalebox{0.8}{MC@NLO}\xspace}
\newcommand{\abr}[1]{\langle #1\rangle}

\newcommand{\mc}[1]{\mathcal{#1}}
\newcommand{\mr}[1]{\mathrm{#1}}

\newcommand{\bs}[1]{\!\!\!\!}
\usepackage{hyperref}
\hypersetup{
  pdfauthor={Stefan Hoeche, Silvan Kuttimalai, Ye Li},
  pdftitle={Hadronic Final States in DIS at NNLO QCD with Parton Showers},
  pdfkeywords={QCD, DIS, HERA, NNLO, Parton Shower}
}
\usepackage{
  pgf,
  tikz}

\usetikzlibrary{
  arrows,
  trees,
  scopes,
  decorations.text,
  decorations.pathreplacing,
  decorations.pathmorphing,
  decorations.markings,
  positioning,
  calc,
  patterns,
  intersections,
  shapes
}

\tikzset{
  dirac/.style=
  {
    draw=black,
    postaction={decorate},
    decoration=
    {
      markings,
      mark=at position .6 with {\arrow[thick,draw=black]{>}}
    }
  },
  majorana/.style=
  {
    draw=black,
  },
  vector/.style=
  {
    decorate,
    draw=black,
    decoration={snake,amplitude=1.5pt,segment length=3pt}
  },
  scalar/.style=
  {
    densely dashed,
    thick,
    draw=black,
  },
  gluon/.style=
  {
    decorate,
    draw=black,
    decoration={coil,amplitude=1.3pt,segment length=1.5pt}
  },
  vertex/.style=
  {
    on grid,
    draw=black,
    fill=black,
    circle,
    minimum size=1.5pt, 
    inner sep=0pt
  },
}

\begin{document}
\preprint{SLAC-PUB-17319, MCNET-18-24}
\title{Hadronic Final States in DIS at NNLO QCD with Parton Showers}
\author{Stefan~H{\"o}che}
\affiliation{SLAC National Accelerator Laboratory,
  Menlo Park, CA, 94025, USA}
\author{Silvan~Kuttimalai}
\affiliation{SLAC National Accelerator Laboratory,
  Menlo Park, CA, 94025, USA}
\author{Ye~Li}
\affiliation{Fermi National Accelerator Laboratory,
  Batavia, IL, 60510-0500, USA}
\begin{abstract}
We present a parton-shower matched NNLO QCD calculation for
hadronic final state production in Deep Inelastic Scattering. 
The computation is based on the \unlops method and is implemented
in the publicly available event generation framework \Sherpa.
Results are compared to measurements performed by the H1 collaboration.
\end{abstract}
\maketitle

\section{Introduction}
Deep-inelastic lepton-nucleon scattering (DIS) has presented
a formidable challenge to the theoretical high-energy physics
community for some time. On the one hand, the reaction provides
an extremely precise probe of the nucleon structure for nearly five
decades~\cite{Devenish:2004pb}. QCD corrections to the structure functions
have been computed through third order in perturbation theory%
~\cite{Zijlstra:1992qd,Zijlstra:1992kj,Moch:1999eb,Vermaseren:2005qc,Moch:2008fj}.
On the other hand, the description of quantities like inclusive jet or
di-jet differential cross sections remained difficult, even with
computations performed at NLO QCD~\cite{Mirkes:1995ks,Graudenz:1997gv,Nagy:2001xb}.
Tremendous progress has recently been made with the fully differential
calculation of jet production in DIS at NNLO QCD accuracy~\cite{Abelof:2016pby,
  Currie:2016ytq,Currie:2017tpe}, and with the fully differential calculation
of inclusive DIS at N$^3$LO precision~\cite{Currie:2018fgr}. Several of these results
have been used in experimental analyses and provide a much improved description
of the measurements~\cite{Andreev:2014wwa,Andreev:2016tgi}.

In this publication we provide a Monte-Carlo simulation for the
estimation of parton shower and hadronization effects that will allow
for a yet more precise comparison between theory and experiment. We
achieve this by combining the known NNLO QCD perturbative
results~\cite{Zijlstra:1992qd,Zijlstra:1992kj,Moch:1999eb} with a
parton shower~\cite{Hoche:2015sya} based on dipole
factorization~\cite{Catani:1996vz}. We implement this simulation in
the general purpose event generator
\Sherpa~\cite{Gleisberg:2003xi,Gleisberg:2008ta}.

\begin{figure}[h]\vskip -1.5mm
  \subfigure[]{
  \begin{tikzpicture}
    \node [vertex] (v)   at (0,0) {};
    
    \node (qi)  at ($(v) +(1,+0.2)$) {};
    \node (qo)  at ($(v) +(1,-0.2)$) {};
    
    \draw [dirac] (v) to [out=+90,in=180, looseness=0.5] (qi) --++ (.3,0);
    \draw         (v) to [out=-90,in=180, looseness=0.5] (qo) --++ (.3,0) node [right, circle,draw=black, inner sep=2] {P};
    
    \node [vertex] (q) at ($(v)+(-1.2,0)$) {};
    \draw [vector] (v) --(q) node [midway, above] {$Q^2$};
    
    \node [inner sep=2] (ei) at ($(q)+(+45+180:1)$) {$k$};
    \node [inner sep=2] (eo) at ($(q)+(-45+180:1)$) {$k^\prime$};
    
    \draw [dirac] (ei) -- (q);
    \draw [dirac] (q)  -- (eo);
  \end{tikzpicture}
  \label{fig:dis-born}
  }\hskip 2.5cm
  \subfigure[]{
  \begin{tikzpicture}
    \node [vertex] (v)   at (0,0) {};
    \node [vertex] (vq)  at ($(v)  +(1,0)$) {};
    
    \draw (v) --++ (0,1) node [near end, right] {$p_T$};
    \draw [dirac] (vq) -- (v);
    \draw ($(vq)+(0,-1)$) -- (vq);
    
    \draw [gluon] (vq) --++ (1,0) node [right, circle, draw=black, inner sep=2] {P};
    
    \node [vertex] (q) at ($(v)+(-1.2,0)$) {};
    \draw [vector] (v) --(q) node [midway, above] {$Q^2$};
    
    \node [inner sep=2] (ei) at ($(q)+(+45+180:1)$) {$k$};
    \node [inner sep=2] (eo) at ($(q)+(-45+180:1)$) {$k^\prime$};
    
    \draw [dirac] (ei) -- (q);
    \draw [dirac] (q)  -- (eo);
  \end{tikzpicture}
  \label{fig:qcd-compton}
  }\vskip -1.5mm
  \caption{Sketch of kinematics in the Breit frame for
    DIS at Born level \subref{fig:dis-born}
    and for the QCD Compton process \subref{fig:qcd-compton}.
    \label{fig:kinematics}}
\end{figure}
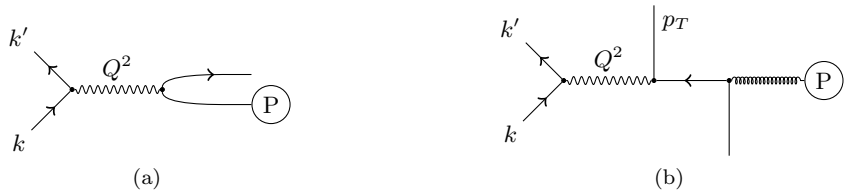

The leading-order QCD configuration corresponding to the DIS process
is shown in Fig.~\ref{fig:dis-born}. The kinematics are typically parametrized
in terms of the exchanged boson's virtuality $Q^2=-q^2=(k-k')^2$
and the Bj{\o}rken variable $x=Q^2/(2\, q P)$.
The jet reconstruction is performed in the Breit frame,
which is defined by the condition $\vec{q}+2x\vec{P}=\vec{0}$.
At the leading order, a single final-state parton emerges,
which carries zero transverse momentum.
The kinematical variables $x$ and $Q^2$ can in principle
be inferred from the incoming and outgoing lepton momenta
$k$ and $k^\prime$ alone, and the only relevant scale in the
problem is $Q^2$. A measurement of jet production in DIS, however,
introduces additional scales of the order of the jet transverse momenta and
leads to significant challenges in the description of the corresponding
final states. The problems are related to possibly inverted scale hierarchies
between $Q^2$ and $p_{T}^2$, the (squared) transverse energy of the
jet(s) in the Breit frame. While the proton is probed at $Q^2$ when
considering inclusive DIS, the relevant hardness scale for the QCD
Compton process leading to the production of a hard jet is of the
order of $Q^2+p_{T}^2$, see the sketch in Fig.~\ref{fig:qcd-compton}. 
In the extreme case of two resolved hard jets and small $Q^2$, one must even
picture the core reaction as a $2\to 2$ pure QCD process, followed by
the initial-state branching of a virtual photon into quarks,
with a relevant hardness scale of $H_T^2$, the total transverse
hadronic energy in the Breit frame.
The large available phase space and the increased scales lead to
significantly enhanced multi-jet cross sections in the region
$p_{T}^2/Q^2\gg 1$. Similar effects were also observed in other
reactions~\cite{Rubin:2010xp}. The problem was addressed in~\cite{Carli:2009cg},
using multi-jet merging.
However, the non-unitary merging technique employed in the simulation
made it difficult to simultaneously predict exclusive quantities
like jet cross sections and inclusive quantities like the structure functions.
In the approach presented here this problem is solved by means of
a unitary merging method. Higher-order radiative effects are taken
into account to a good approximation by including the complete
fixed-order NNLO QCD corrections to inclusive DIS and by choosing
an appropriate scale~\cite{Currie:2017tpe}.

This paper is organized as follows: Section~\ref{sec:methods} presents
an introduction to the \unlops matching technique and discusses
the projection-to-Born method used in our study.
Section~\ref{sec:results} presents the numerical validation, assessment
of theoretical uncertainties and comparison to experimental data.
An outlook is given in Sec.~\ref{sec:conclusions}.

\section{Computational setup}
\label{sec:methods}

The starting point of our simulation is a fully differential
calculation of the inclusive DIS process at NNLO QCD using the
projection-to-Born method~\cite{Cacciari:2015jma}. This technique relies
on a map $F_b$ that uniquely assigns any given flavor and momentum
configuration in the one emission phase space $\Phi_1$ and in the
double-real emission phase space $\Phi_2$ to a point in the Born
phase space $\Phi_0$. Given such a map, we define the Born-differential
NNLO cross section as
\begin{align}
  \begin{split}
  \bar{\bar{\mr{B}}}(\Phi_0)
  = \;&\mr{B}_0(\Phi_0) + \mr{V}_0(\Phi_0) + \mr{VV}_0(\Phi_0) \\
    &+\int\dif\Phi_1\left( \mr{B}_1(\Phi_1) + \mr{V}_1(\Phi_1)\right)\delta^{(2)}\!\left(F_B(\Phi_1) - \Phi_0\right) \\
    &+\int\dif\Phi_2\; \mr{B}_2(\Phi_2)\,\delta^{(2)}\!\left(F_B(\Phi_2) - \Phi_0\right) \label{eq:dsignnlo-def}
  \end{split}
\end{align}
In this context, $\mr{B}_n$ are the Born differential cross sections
for DIS plus $n$ partons, and $\mr{V}_n$ and $\mr{VV}_0$ are the
corresponding UV renormalized virtual and double virtual corrections,
including the appropriate collinear mass factorization counterterms.
This cross section is free of divergences if $F_B$ is an infrared safe
observable, i.e.\ if it maps a Born phase space point supplemented
by infinitely soft gluons and/or collinear parton branchings to the
same Born phase space point. In DIS, the construction of $F_B$ is
straightforward: We require that the mapping preserve the lepton
momenta. This choice uniquely determines the kinematics of
the corresponding Born configuration, where the momentum of the incoming
QCD parton is then set to $p=xP$ and the momentum of the outgoing QCD
parton is given by $p+q$.

In terms of the Born-differential NNLO cross section, any infrared-safe
observable $O$ can now be calculated as follows:
\begin{equation}
  \begin{split}
  \abr{O}^{\rm(NNLO)}
  =\; &\int\dif\Phi_0\;\bar{\bar{\mr{B}}}(\Phi_0)\,O(\Phi_0) \\
  &+\int\dif\Phi_1 \left(\mr{B}_1(\Phi_1) + \mr{V}_1(\Phi_1)\right)\Big[\,O(\Phi_1) - O(F_B(\Phi_1))\,\Big] \\
  &+\int\dif\Phi_2\; \mr{B}_2(\Phi_2)\Big[\,O(\Phi_2) - O(F_B(\Phi_2))\,\Big] \label{eq:ptb}
  \end{split}
\end{equation}
While the first line of Eq.~\eqref{eq:ptb} generates the observable
dependence correctly in the Born phase space, the second and third
line correct for it's dependence in the single and double emission
phase space. They are generated by events in
the single and double emission phase space with the appropriate
flavor and momentum configurations (corresponding to the $O(\Phi_1)$
and $O(\Phi_2)$ terms). For each event, a duplicate event with
inverted weight and Born-projected flavor-kinematics structure is
added (corresponding to the $O(F_B(\Phi_1))$ and
$O(F_B(\Phi_2))$ terms). Since the events in the single and double
emission phase space correspond to a regular NLO calculation of the
jet-associated Born process, any of the well established techniques
for the computation of virtual corrections and infrared-subtraction
at NLO can be used. In our work, we employ the \BlackHat library
\cite{Berger:2008sj,Berger:2009ep,Berger:2010vm} for the computation
of the virtual corrections and Catani-Seymour dipole subtraction
\cite{Catani:1996vz} as implemented in the matrix element generator
\Amegic~\cite{Krauss:2001iv,Gleisberg:2007md}. Our implementation of the
generic NNLO corrections in Eq.~\eqref{eq:dsignnlo-def} is based
on the two-loop DIS structure functions available in the literature
\cite{Zijlstra:1992qd,Zijlstra:1992kj,Moch:1999eb}.

We match the fixed-order computation in the projection-to-Born method
to a parton shower using the \unlops algorithm~\cite{Lonnblad:2012ix,Hoeche:2014aia}.
The effect of additional emissions generated in the parton shower approach
can be described using a generating functional, which is recursively defined for an
$n$-parton final state and the observable $O$ as
\begin{equation}\label{eq:ps_functional}
  \mc{F}_n(t_n,O;\Phi_n)=\Pi_n(t_c,t_n;\Phi_n)\,O(\Phi_n)
  +\int_{t_c}^{t_n}\dif\hat{\Phi}_1\,\mr{K}_n(\Phi_n,\hat{\Phi}_1)\,
  \Pi_n(\hat{t},t_n;\Phi_n)\,\mc{F}_{n+1}(\hat{t},O;\Phi_{n+1})\;,
\end{equation}
where $\hat{t}=t(\hat{\Phi}_1)$, and where the parton-shower
no-branching probability is given by
\begin{equation}\label{eq:ps_sudakov}
  \Pi_n(t,t';\Phi_n)=\exp\left\{-\int_t^{t'}\dif\hat{\Phi}_1\,\mr{K_n}(\Phi_n,\hat{\Phi}_1)\right\}\;.
\end{equation}
Here, $\dif\hat{\Phi}_1$ is the differential one-emission phase space, which is
parametrized in terms of the evolution and splitting variables $t$ and $z$
as $\dif\hat\Phi_1=\dif t\,\dif z\,\dif\phi/(2\pi)J(t,z,\phi)$,
with $J$ a possible Jacobian factor. The starting scale of the evolution is given
by $t_n$, while $t_c$ denotes the cutoff scale. $\mr{K}_n$ is the evolution kernel
for the $n$-parton state. In the case of DGLAP evolution of the DIS process,
it can be written as~\cite{Webber:1986mc,Buckley:2011ms}
\begin{equation}
  \mr{K}_n(\Phi_n,\hat{\Phi}_1)=
  \sum_{b=q,g}\frac{\alpha_s}{2\pi}\,P_{ba}(z)
  \frac{f_b(x/z,t)}{z\,f_{a}(x,t)}\,\Theta(z-x)+
  \sum_{i=1}^{n_{\rm out}}\,\sum_{b=q,g}\frac{\alpha_s}{2\pi}\,P_{a_ib}(z)\;,
\end{equation}
where the first term corresponds to initial-state radiation and the second term
to final-state radiation. The matching to a fixed-order higher-order calculation
is achieved by exploiting the factorization of tree-level matrix elements, which
implies schematically that $\mr{B}_{n+1}\to\mr{B}_n\mr{K}_n$ in the soft or collinear
limit. Note that in processes with a more complicated color structure or external
gluons, spin and color correlations between the underlying Born configuration,
$\mr{B}_n$, and the splitting kernels in $\mr{K}_n$ must be taken into account.
In our simulation of the DIS di-jet topologies, these correlations are included by means
of an NLO matching in the \mcatnlo method~\cite{Hoeche:2011fd,Hoeche:2012fm}.

The NNLO matching in the \unlops method proceeds in two steps. In
order to reproduce the logarithmic coefficients of the parton shower
resummation, the real emission terms in the fixed-order calculation
are reweighted. The nominal accuracy of the fixed-order NNLO
calculation is then restored by subtracting the fixed-order expansion
of the reweighted result to the second order in the strong coupling
such as to remove the overlap with the exact NNLO result. In
Eq.~\eqref{eq:nnlo_ps} we quote only the final formula. The details of
the matching procedure are given
in~\cite{Lonnblad:2012ix,Hoeche:2014aia}. The only modifications of
the original method that lead to Eq.~\eqref{eq:nnlo_ps} are due to our
NNLO fixed-order input being computed in the projection-to-Born
method, rather than the $q_T$-cutoff technique.

We start with the one-jet differential NLO cross sections for standard
and hard events as defined in the \mcatnlo method~\cite{Frixione:2002ik,Hoeche:2011fd},
\begin{equation}\label{eq:def_bbar_h_mcnlo}
  \begin{split}
  \tilde{\mr{B}}_1(\Phi_1)=&\;\mr{B}_1(\Phi_1)+\tilde{\mr{V}}_1(\Phi_1)+\mr{I}_1(\Phi_1)
  -\int_{t_c}\dif\hat{\Phi}_1\,\mr{S}_1(\Phi_1,\hat{\Phi}_1)\,\Theta(t_2(\hat{\Phi}_1)-t_1(\Phi_1))\;,\\
  \mr{H}_1(\Phi_2)=&\;\mr{B}_2(\Phi_2)-\mr{S}_1(\Phi_2)\,\Theta(t_1(\Phi_2)-t_2(\Phi_2))\;.
  \end{split}
\end{equation}
The generating functional for matching the one-jet process at NLO is given
in terms of dipole subtraction terms, $\mr{S}_1$, and the underlying Born
differential cross sections, $\mr{B}_1$, as
\begin{equation}
  \tilde{\mc{F}}_1(t,O;\Phi_1)=\tilde{\Pi}_1(t_c,t_1;\Phi_1)\,O(\Phi_1)
      +\int_{t_c}\dif\hat{\Phi}_1\frac{\mr{S}_1(\Phi_1,\hat{\Phi}_1)}{\mr{B}_1(\Phi_1)}\,
      \tilde{\Pi}_1(\hat{t},t_1;\Phi_1)\,\mc{F}_2(\hat{t},O;\Phi_2)\;.
\end{equation}
The no-emission probability is defined as in Eq.~\eqref{eq:ps_sudakov}
with $\mr{K}_1\to\mr{S}_1/\mr{B}_1$.
Events are generated above the parton-shower cutoff scale, $t_c$,
below which the DIS process is considered to be inclusive. We introduce
a regular and an exceptional part of the \mcatnlo hard remainder function
\begin{align}\label{eq:unordered_hevents}
  \mr{H}_1^{\mr{R}}(\Phi_2)
  &= \mr{H}_1(\Phi_2)
     \Theta\left(t_1-t_2\right)
     \Theta\left(t_2-t_c\right)\;,
  &\mr{H}_1^{\mr{E}}(\Phi_2)
  &= \mr{H}_1(\Phi_2)-\mr{H}_1^{\mr{R}}(\Phi_2)\;.
\end{align}
Exceptional contributions appear in regions of phase space for which no ordered
parton shower history can be identified. The prime example are configurations
where the transverse momenta of jets in the Breit frame are much larger than $Q^2$,
cf.\ the sketch in Fig.~\ref{fig:qcd-compton}.

The final \unlops matching formula at NNLO accuracy reads
\begin{equation}\label{eq:nnlo_ps}
  \begin{split}
    &\abr{O}^\mr{(UNLOPS)}=\;
    \int\dif\Phi_0\,\bar{\bar{\mr{B}}}_0(\Phi_0)\,O(\Phi_0)\\
    &\quad+\int_{t_c}\dif\Phi_1\,
    \Pi_0(t_1,\mu_Q^2)\Big(w_1(\Phi_1)+w_1^{(1)}(\Phi_1)+\Pi_0^\mr{(1)}(t_1,\mu_Q^2)\Big)
    \,\mr{B}_1(\Phi_1)\,\Big[\tilde{\mc{F}}_1(t_1,O;\Phi_1)-O(\Phi_0)\Big]\\
    &\quad+\int_{t_c}\dif\Phi_1\,
    \Pi_0(t_1,\mu_Q^2)\,\tilde{\mr{B}}_1^{\rm{R}}(\Phi_1)\,\,
    \Big[\tilde{\mc{F}}_1(t_1,O;\Phi_1)-O(\Phi_0)\Big]\\
    &\quad+\int_{t_c}\dif\Phi_2\,
    \Pi_0(t_1,\mu_Q^2)\,\mr{H}_1^{\mr{R}}(\Phi_2)\,\Big[\mc{F}_2(t_2,O;\Phi_2)-O(\Phi_0)\Big]
    +\int_{t_c}\dif\Phi_2\,
      \,\mr{H}_1^{\mr{E}}(\Phi_2)\,\mc{F}_2(t_2,O;\Phi_2)\;.
  \end{split}
\end{equation}
We have defined $\tilde{\mr{B}}_1^{\rm{R}}=\tilde{\mr{B}}_1-\mr{B}_1$ and
introduced the \unlops matching weight, $w_1$, which is given by~\cite{Lonnblad:2012ix}
\begin{equation}\label{eq:unlops_weight}
  w_1(\Phi_1)=\frac{\alpha_s(t_1)}{\alpha_s(\mu_R^2)}\,
  \frac{f_a(x_a,t_1)}{f_a(x_a,\mu_F^2)}
  \frac{f_{a'}(x_{a'},\mu_F^2)}{f_{a'}(x_{a'},t_1)}\,
  \bigg[1+\delta_{aa'}\frac{\alpha_s(t_1)}{2\pi}\,K\bigg]\;.
\end{equation}
$f_a(x_a)$ and $f_{a'}(x_{a'})$ denote the PDFs associated with the external
and intermediate parton, respectively, and the constant
$K=(67/18-\pi^2/6)\,C_A-10/9\,T_R\,n_f$ is the 2-loop cusp anomalous dimension,
which restores the physical coupling in soft-gluon emissions~\cite{Catani:1990rr,Banfi:2018mcq}.
The subtraction terms for the no-branching probability of the parton shower,
and for the weight $w_1$, are given by 
\begin{equation}\label{eq:unnlops_weight_subterms}
  \begin{split}
    \Pi_0^{(1)}(t,t')=&\;\int_t^{t'}\dif\hat{\Phi}_1\,
    \frac{\alpha_s(\mu_R^2)}{\alpha_s(\hat{t})}\,
    \bigg[1+\delta_{aa'}\frac{\alpha_s(t_1)}{2\pi}\,K\bigg]^{-1}\,
    \mr{K}_1(\Phi_1,\hat{\Phi}_1)\\
    w_1^{(1)}(\Phi_1)=&\;\frac{\alpha_s(\mu_R^2)}{2\pi}\Bigg[\,\beta_0\ln\frac{b_{a\!a'}\,t_1}{\mu_R^2}
    -\ln\frac{t_1}{\mu_F^2}\sum_c\bigg(\int_x^1\frac{\dif z}{z}P_{ca}(z)\,
    \frac{f_c(x/z,\mu_F^2)}{f_a(x,\mu_F^2)}
    -\int_{x'}^1\frac{\dif z}{z}P_{c{a'}}(z)\,
    \frac{f_c(x'/z,\mu_F^2)}{f_{a'}(x',\mu_F^2)}\bigg)\Bigg]\;,
  \end{split}
\end{equation}
where $b_{aa'}=\exp\{-\delta_{aa'}K/\beta_0\}$.
Finally, the resummation scale in Eq.~\eqref{eq:nnlo_ps} is given by $\mu_Q^2=\max(t_1,Q^2)$.

The importance of scale choices in higher-order perturbative QCD calculations
has been in the focus of interest recently~\cite{Anger:2017nkq,Currie:2018xkj}.
In order to reflect the dynamics of the DIS di-jet and tri-jet processes
in the high transverse momentum region, we select a scale similar to the
one proposed in~\cite{Currie:2017tpe}. Instead of the jet transverse
momenta we employ the total transverse hadronic energy in the Breit frame,
$H_{T}$. The central renormalization and factorization
scale in our fixed-order calculations is then given by
\begin{equation}\label{eq:fo_scale}
  \mu_{R/F}^2=\frac{Q^2+(H_{T}/2)^2}{2}
\end{equation}
Equation~\eqref{eq:fo_scale} smoothly interpolates between the regions
of normal scale hierarchies, where $Q$ is larger than the transverse
momenta of jets in the Breit frame, and the regions of inverted scale
hierarchies, where the transverse momenta are much larger than $Q$.
This corresponds to the two situations sketched in Fig.~\ref{fig:kinematics}.
Our scale choice Eq.~\eqref{eq:fo_scale} effectively selects largest scale
in the reaction in both cases.

\section{Results}
\label{sec:results}
In this section we present numerical results of our calculation,
quantitative assessments of the theoretical uncertainties, and
comparisons to experimental data. We use the publicly available
event generation framework \Sherpa~\cite{Gleisberg:2003xi,Gleisberg:2008ta},
modified to include the changes described in Sec.~\ref{sec:methods}.
Parton showers are generated using the \Dire model~\cite{Hoche:2015sya}.
Comparing the radiation pattern of \Dire in higher-multiplicity events
to fixed-order predictions we find that the simulation can be improved
by redefining the evolution variables for dipoles with initial-state emitter
and final state spectator or vice versa as $t=Q^2\,u\,(1-z)/z$,
as opposed to the definition $t=Q^2\,u\,(1-z)$ given in~\cite{Hoche:2015sya}.
We use the CT14nnlo PDF set~\cite{Dulat:2015mca} and choose the strong coupling accordingly.
Analyses of the simulated events are performed with the help of
\Rivet~\cite{Buckley:2008vh,Buckley:2010ar}.

\begin{figure}[t]
  \centering
  \begin{minipage}{0.4\textwidth}
    \includegraphics[width=\textwidth]{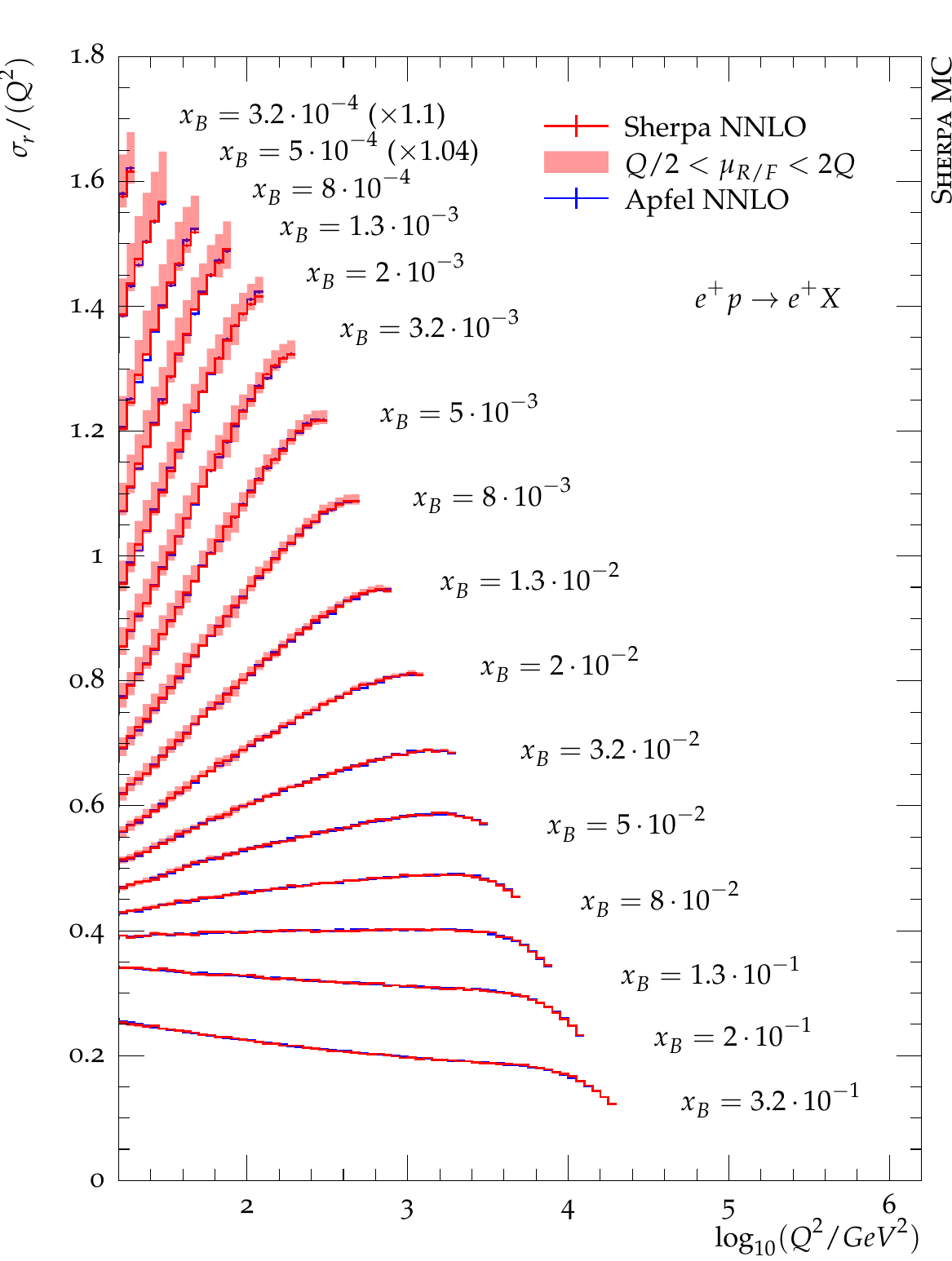}
  \end{minipage}\hskip 1.5cm
  \begin{minipage}{0.4\textwidth}
    \includegraphics[width=\textwidth]{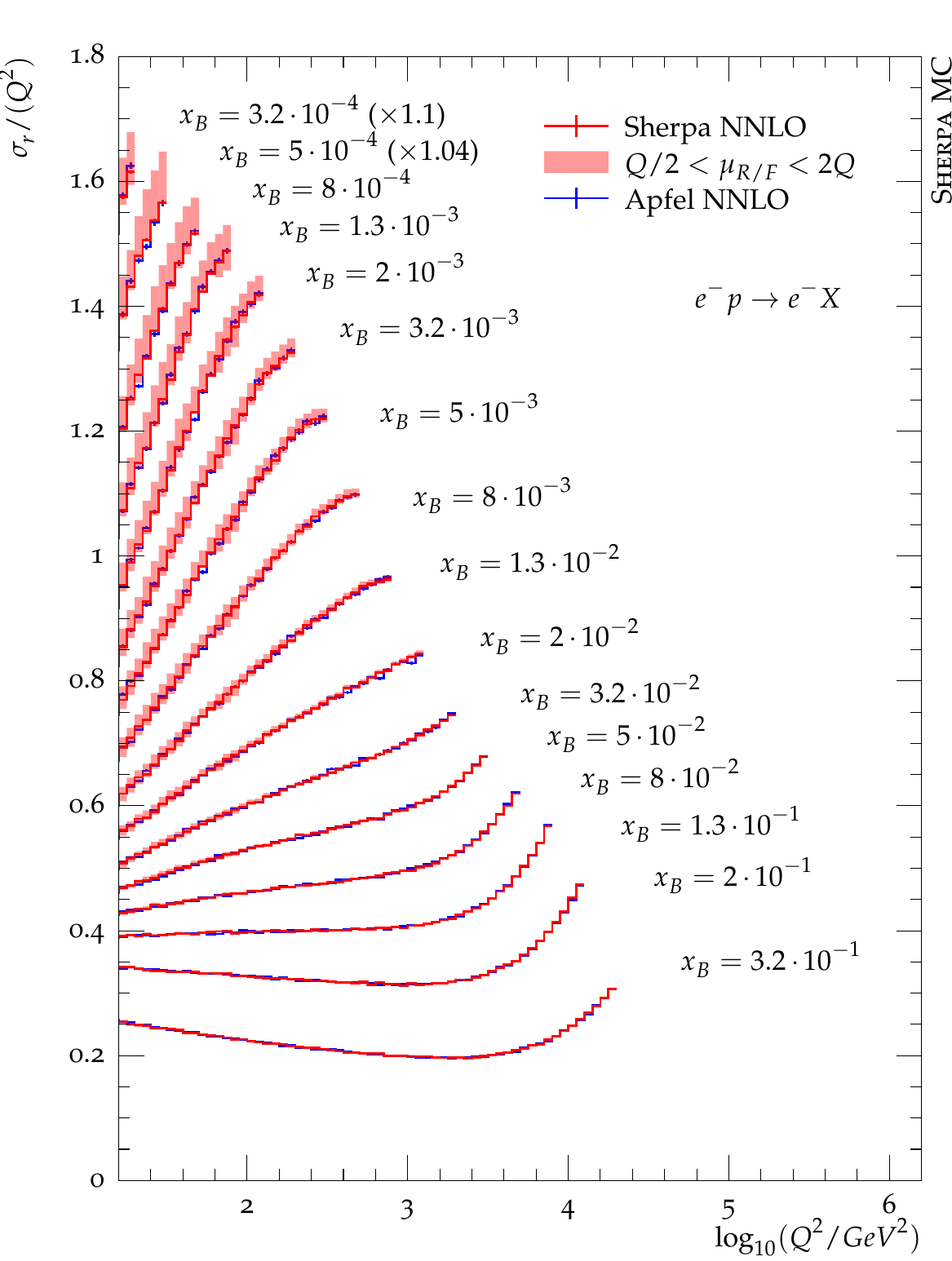}
  \end{minipage}
  \caption{Comparison of NNLO cross sections, differential in $x$ and $Q^2$,
    between our dedicated implementation in \Sherpa and the publicly available
    \Apfel library~\cite{Bertone:2017gds}.\label{fig:nnlo_xs_comparison}}
\end{figure}

In order to validate our implementation of the Born differential NNLO
cross section, we compare fixed-order predictions for the reduced
cross section to results obtained with the publicly available \Apfel
library~\cite{Bertone:2017gds}. The reduced cross section is defined as
\begin{align}
  \sigma_r = \frac{{\dif\,}^2\sigma}{\dif x\dif Q^2}
  \frac{Q^4 x}{2\pi\alpha^2 Y_+}\label{eq:red-xs}\;,
\end{align}
where $Y_+=(1+(1-y)^2)$ and $y=Q^2/(sx)$ with $s$ the center-of-mass
energy of the collider. As shown in Figure~\ref{fig:nnlo_xs_comparison},
where we show a comparison differentially in $Q^2$ and $x$, we observe good
agreement within the statistical uncertainties. Note that for this cross-check
we set the renormalization and factorization scale to $\mu_{R/F}^2=Q^2$.

\begin{figure}[p]
  \centering
  \begin{minipage}{0.325\textwidth}\vskip 4mm
    \includegraphics[width=\textwidth]{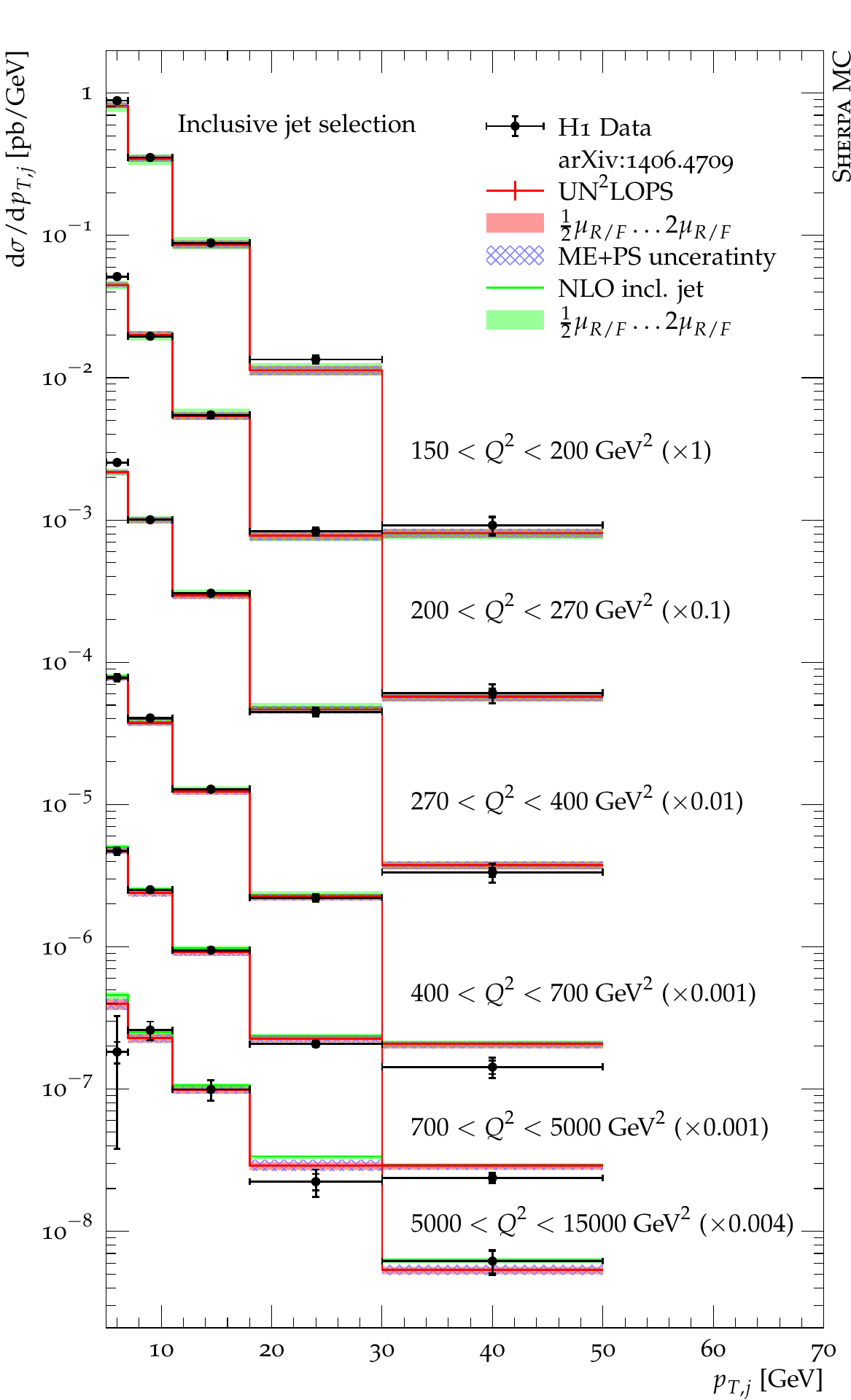}
  \end{minipage}\hfill
  \begin{minipage}{0.325\textwidth}\vskip 4mm
    \includegraphics[width=\textwidth]{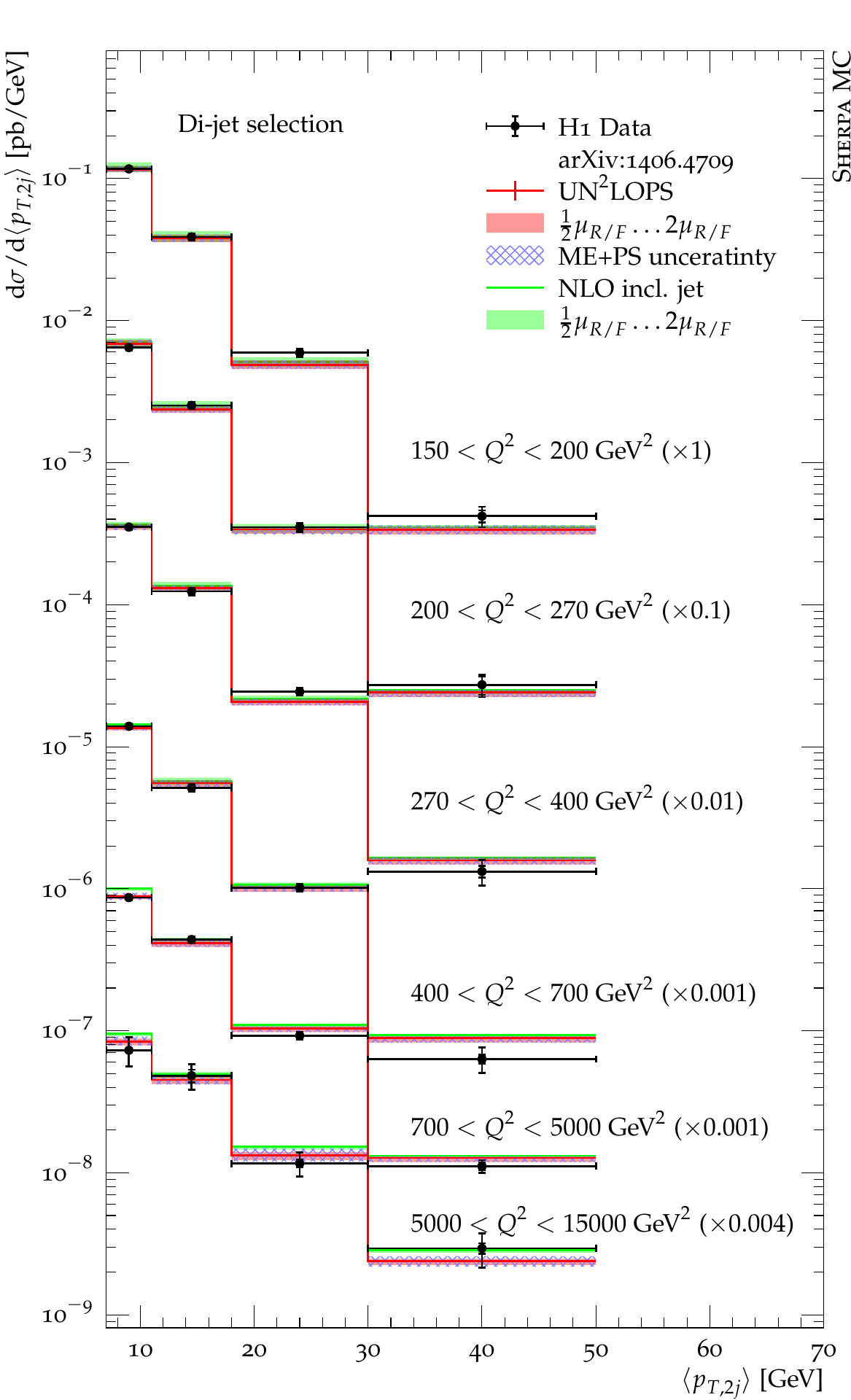}
  \end{minipage}\hfill
  \begin{minipage}{0.325\textwidth}\vskip 4mm
    \includegraphics[width=\textwidth]{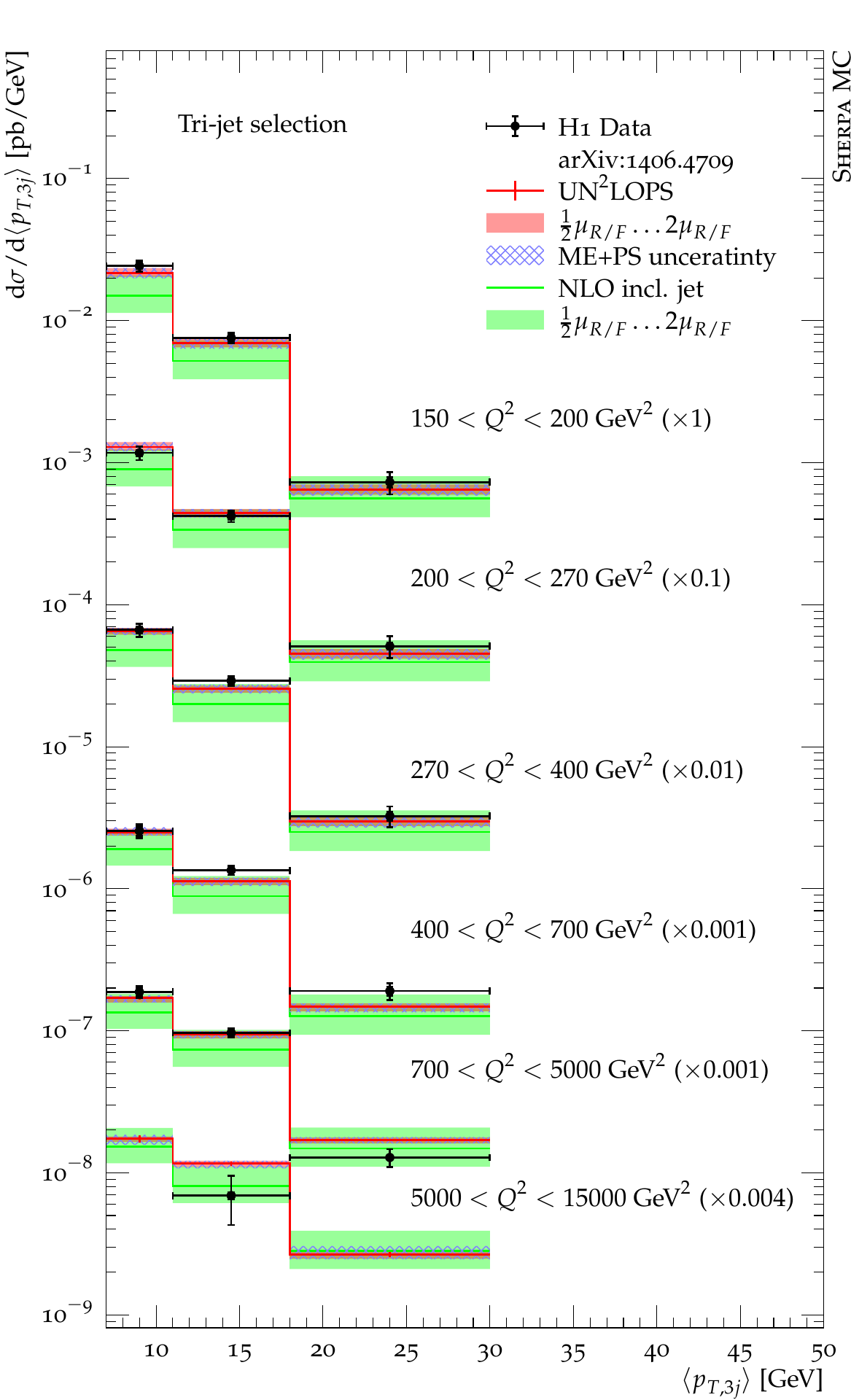}
  \end{minipage}\\
  \begin{minipage}{0.325\textwidth}\vskip 4mm
    \includegraphics[width=\textwidth]{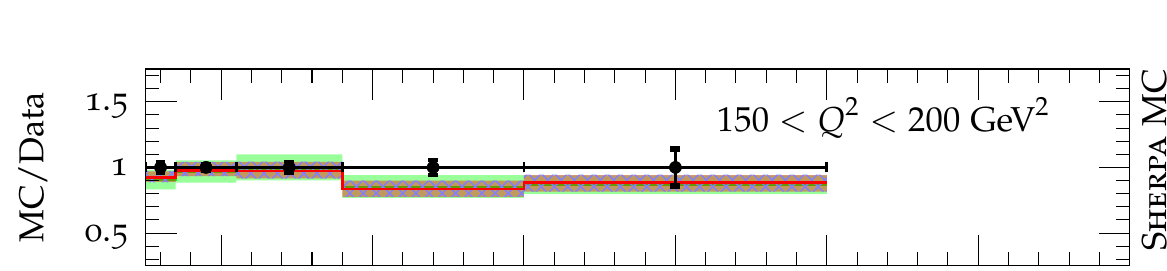}\\[-1mm]
    \includegraphics[width=\textwidth]{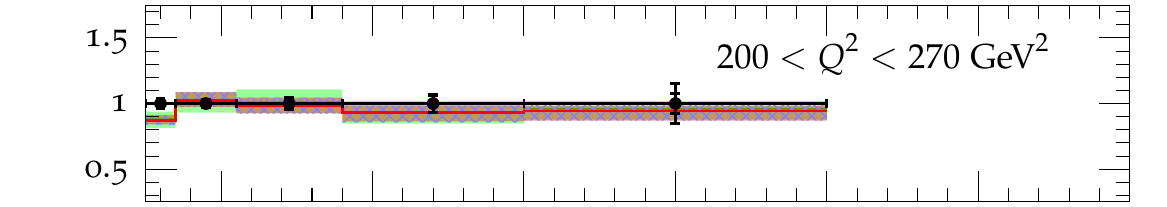}\\[-1mm]
    \includegraphics[width=\textwidth]{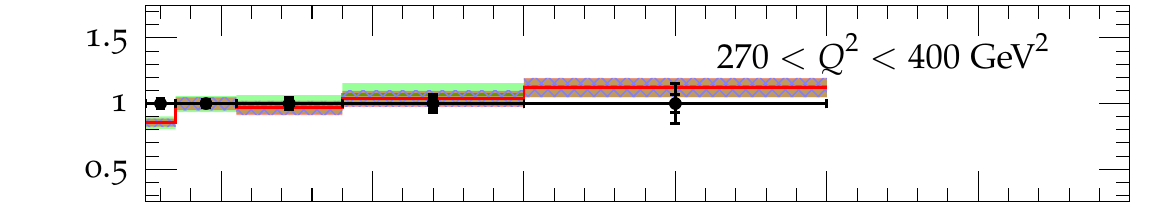}\\[-1mm]
    \includegraphics[width=\textwidth]{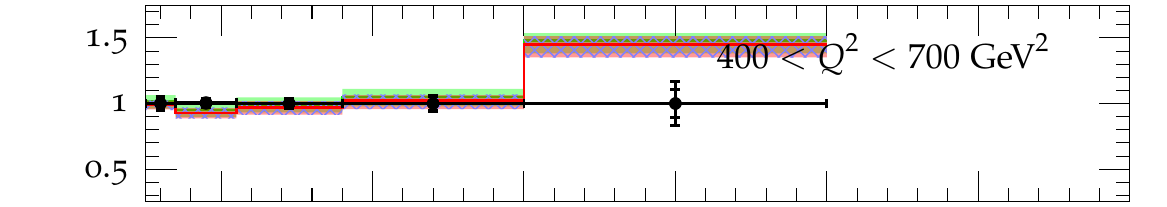}\\[-1mm]
    \includegraphics[width=\textwidth]{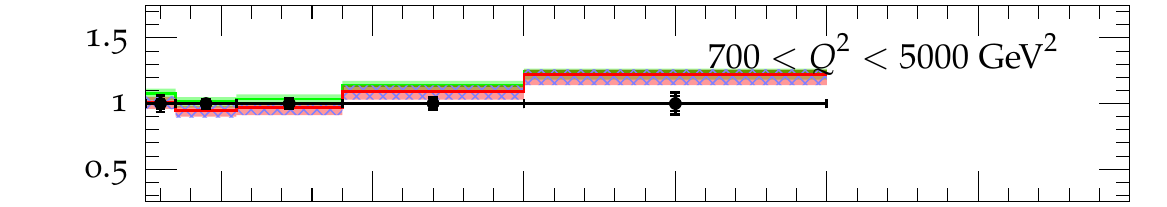}\\[-1mm]
    \includegraphics[width=\textwidth]{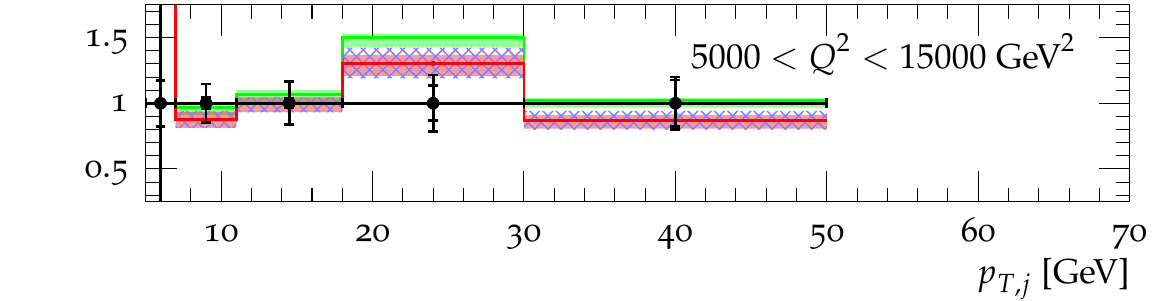}
  \end{minipage}\hfill
  \begin{minipage}{0.325\textwidth}\vskip 4mm
    \includegraphics[width=\textwidth]{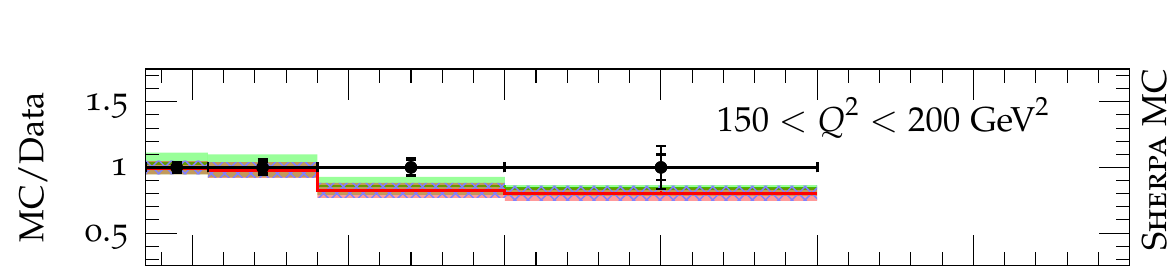}\\[-1mm]
    \includegraphics[width=\textwidth]{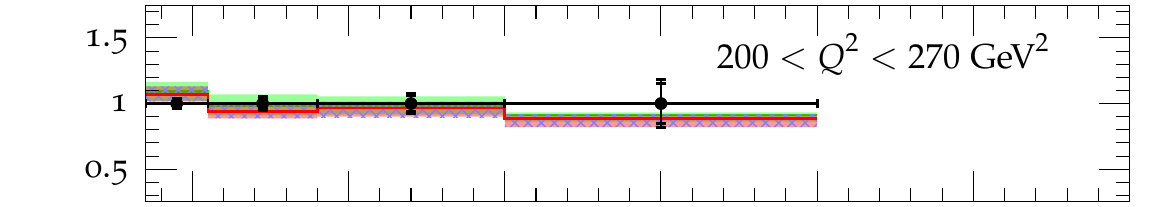}\\[-1mm]
    \includegraphics[width=\textwidth]{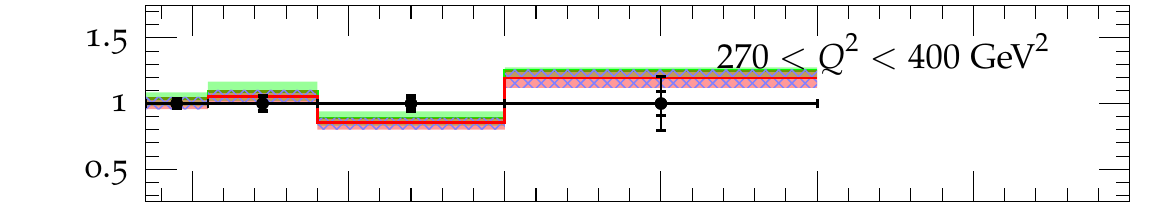}\\[-1mm]
    \includegraphics[width=\textwidth]{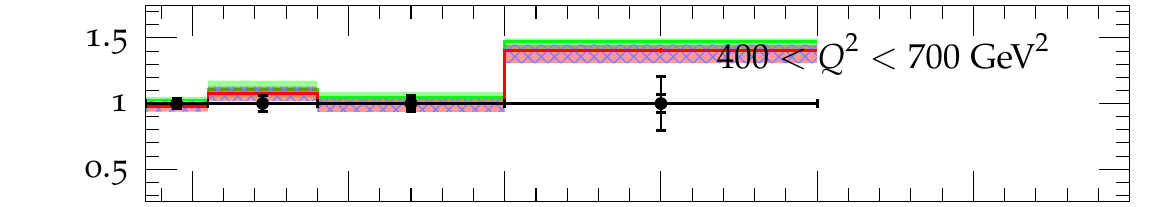}\\[-1mm]
    \includegraphics[width=\textwidth]{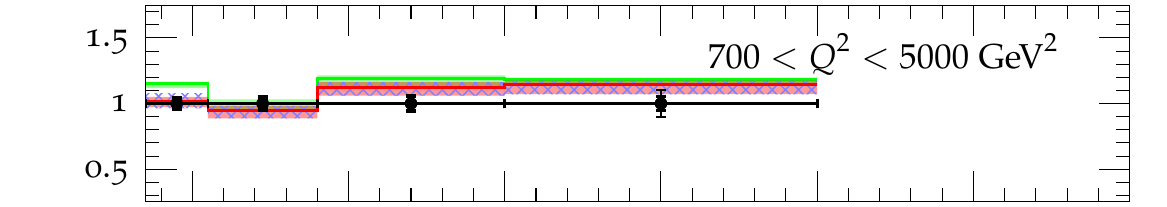}\\[-1mm]
    \includegraphics[width=\textwidth]{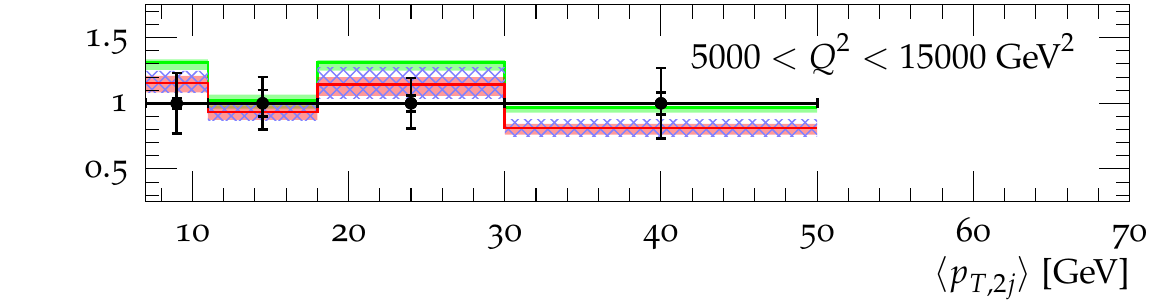}
  \end{minipage}\hfill
  \begin{minipage}{0.325\textwidth}\vskip 4mm
    \includegraphics[width=\textwidth]{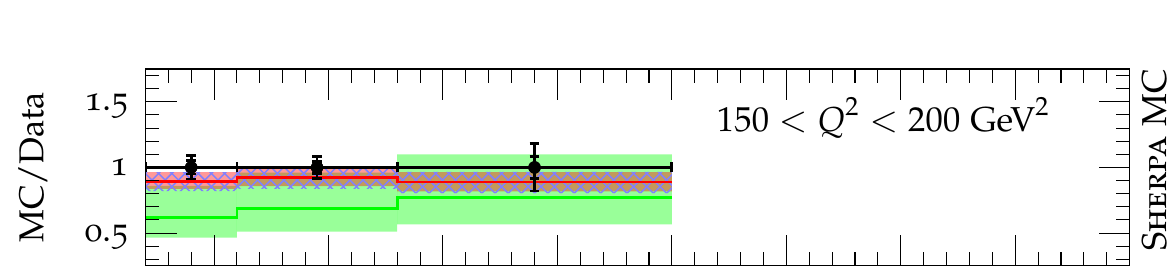}\\[-1mm]
    \includegraphics[width=\textwidth]{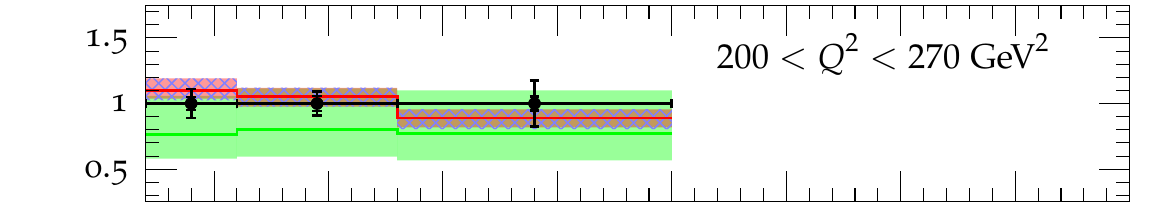}\\[-1mm]
    \includegraphics[width=\textwidth]{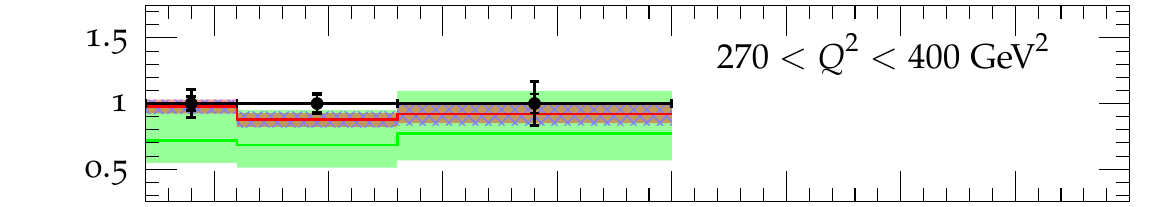}\\[-1mm]
    \includegraphics[width=\textwidth]{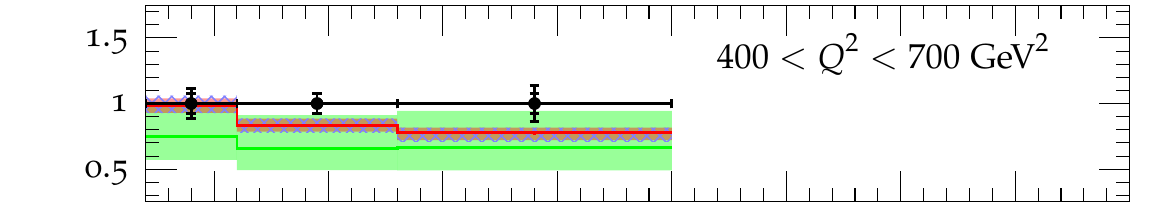}\\[-1mm]
    \includegraphics[width=\textwidth]{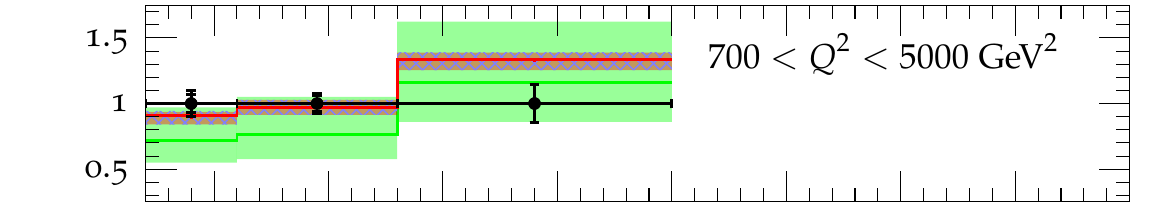}\\[-1mm]
    \includegraphics[width=\textwidth]{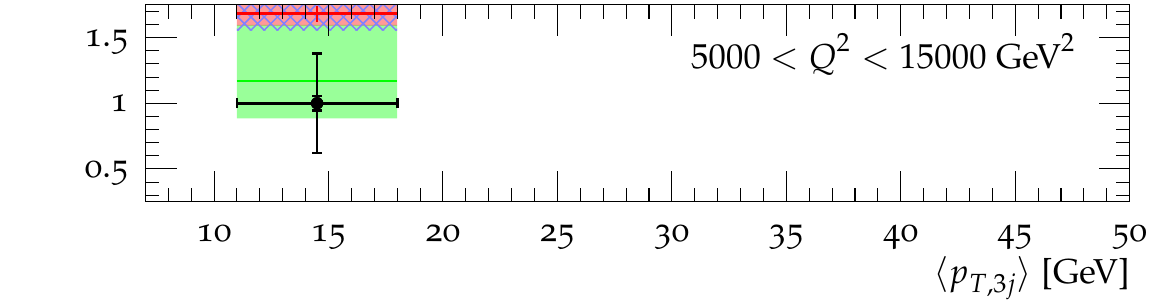}
  \end{minipage}
  \caption{Inclusive jet, di-jet and tri-jet cross section differential in $Q^2$
    as a function of $p_{T,j}$ compared to experimental data from the H1 collaboration
    \cite{Andreev:2014wwa}. We show separate error bars for the reported
    statistical and systematic uncertainties. We compare NLO fixed-order predictions
    corrected for hadronization effects (green) and parton-shower matched
    NNLO predictions at the particle level (red). The light green and
    light red uncertainty bands are obtained from a correlated variation
    of the renormalization and factorization scales. The hatched blue
    uncertainty band combines the fixed-order uncertainties and
    parton-shower uncertainties in quadrature. See the main text for details.
    \label{fig:inclusive_jetpt_highq2}}
\end{figure}

\begin{figure}[p]
  \centering
  \begin{minipage}{0.325\textwidth}\vskip 4mm
    \includegraphics[width=\textwidth]{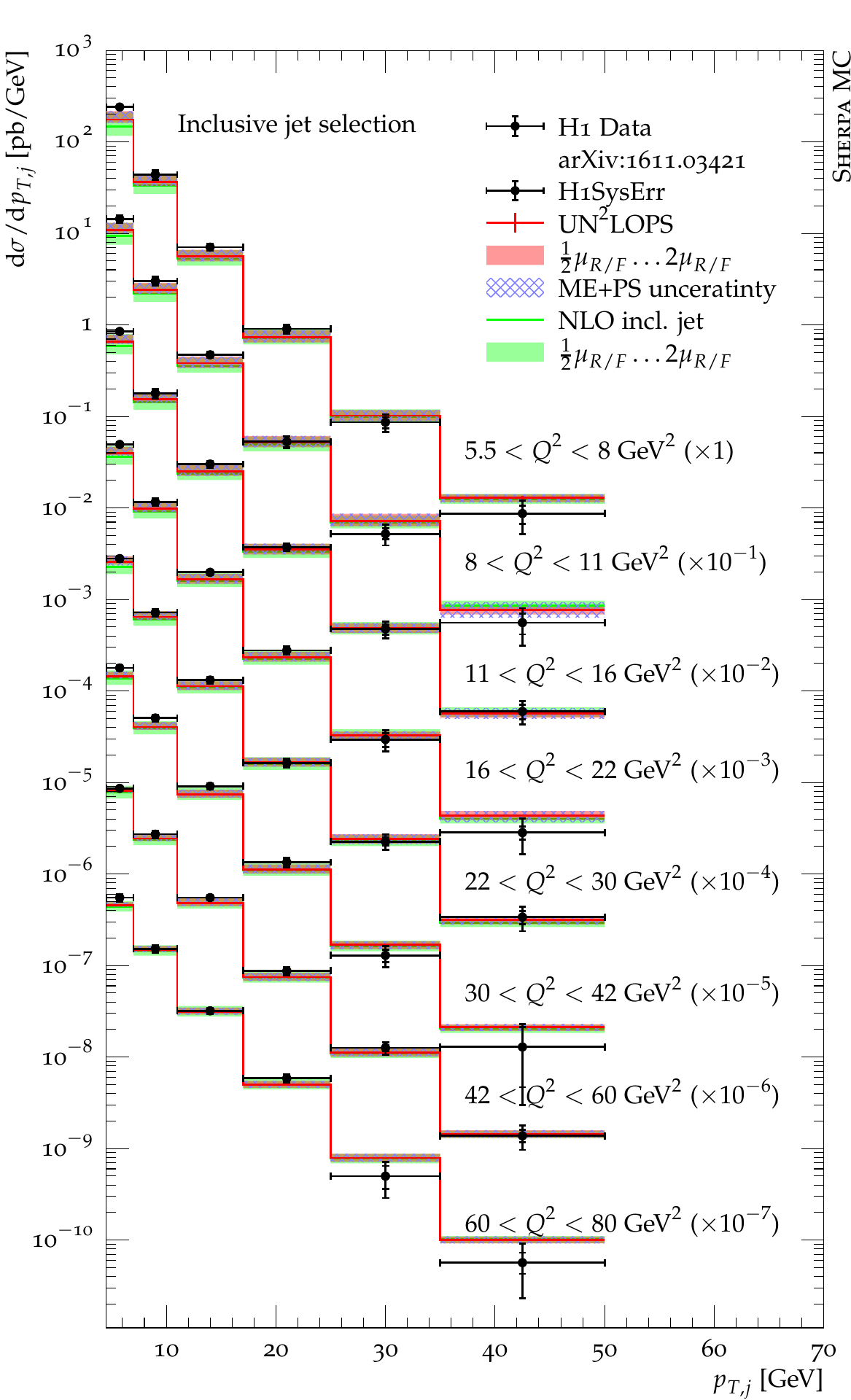}
  \end{minipage}\hfill
  \begin{minipage}{0.325\textwidth}\vskip 4mm
    \includegraphics[width=\textwidth]{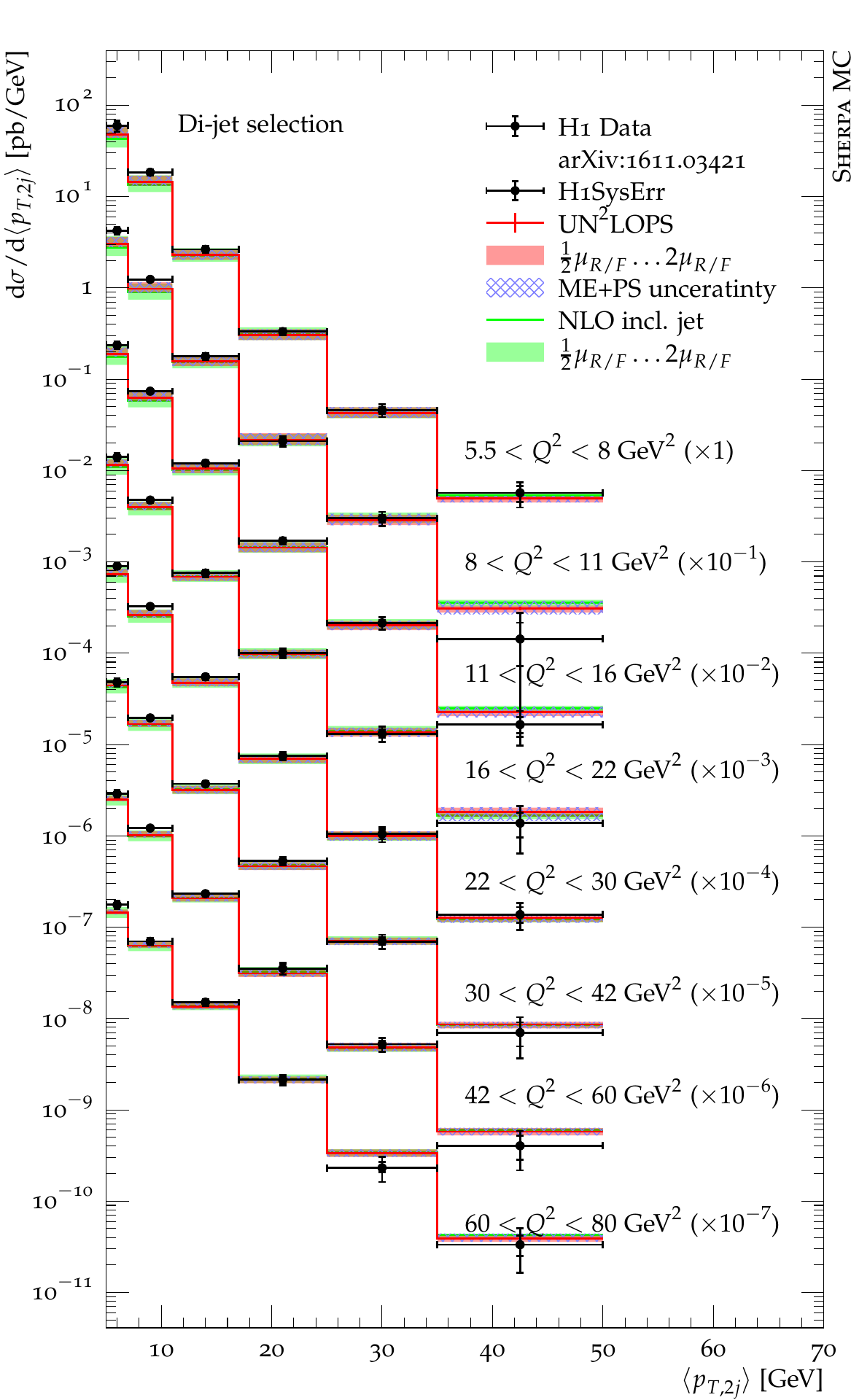}
  \end{minipage}\hfill
  \begin{minipage}{0.325\textwidth}\vskip 4mm
    \includegraphics[width=\textwidth]{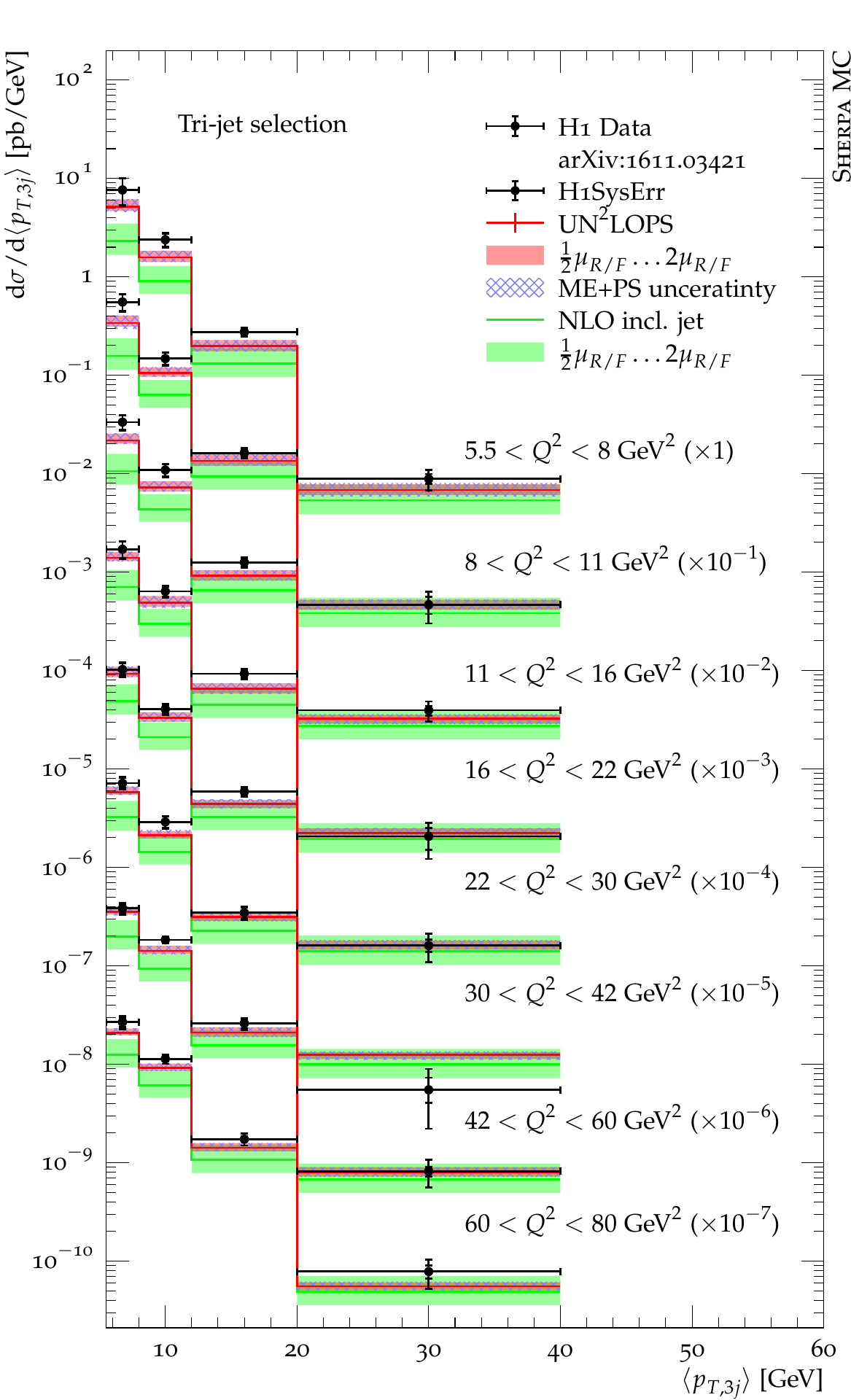}
  \end{minipage}\\
  \begin{minipage}{0.325\textwidth}\vskip 4mm
    \includegraphics[width=\textwidth]{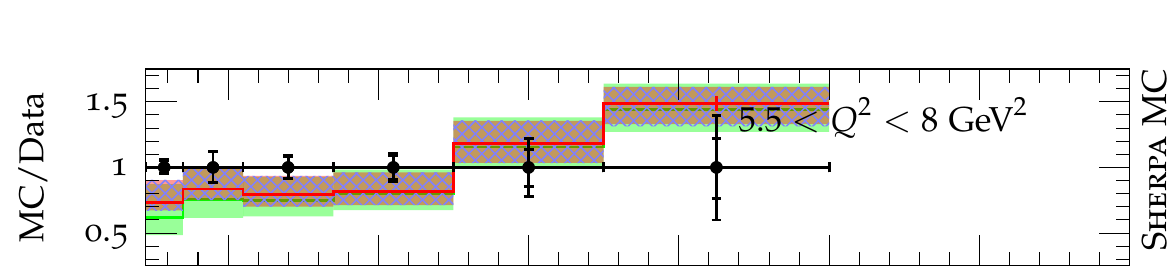}\\[-1mm]
    \includegraphics[width=\textwidth]{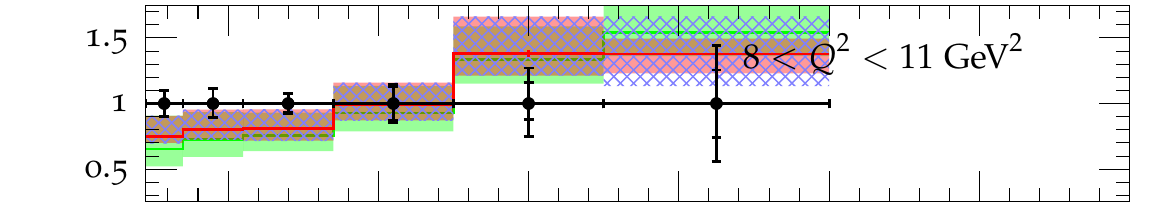}\\[-1mm]
    \includegraphics[width=\textwidth]{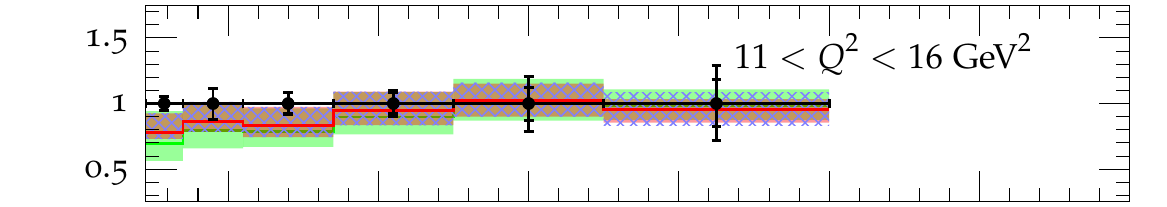}\\[-1mm]
    \includegraphics[width=\textwidth]{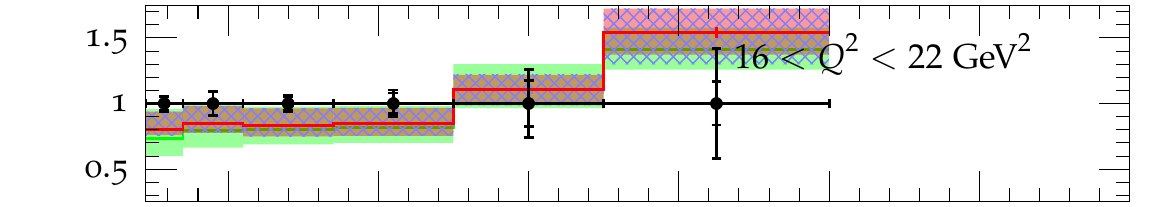}\\[-1mm]
    \includegraphics[width=\textwidth]{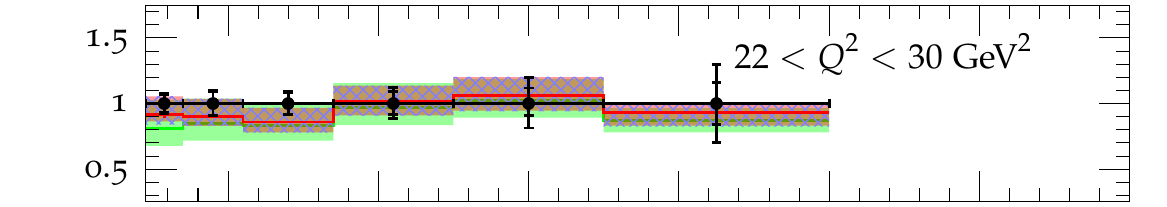}\\[-1mm]
    \includegraphics[width=\textwidth]{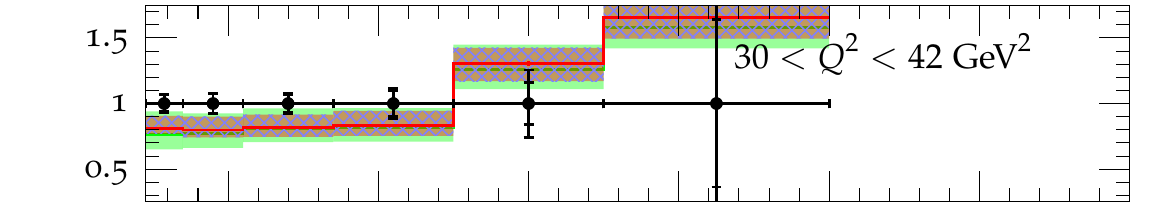}\\[-1mm]
    \includegraphics[width=\textwidth]{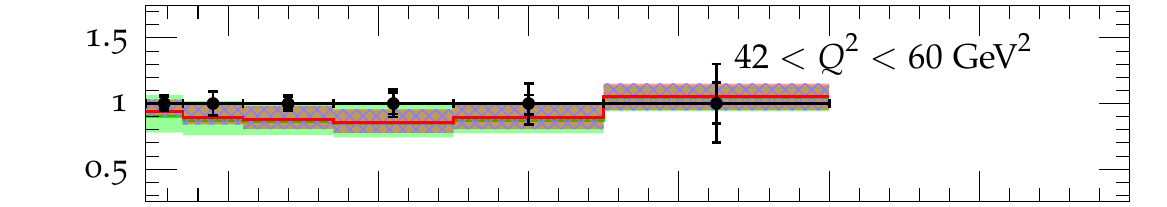}\\[-1mm]
    \includegraphics[width=\textwidth]{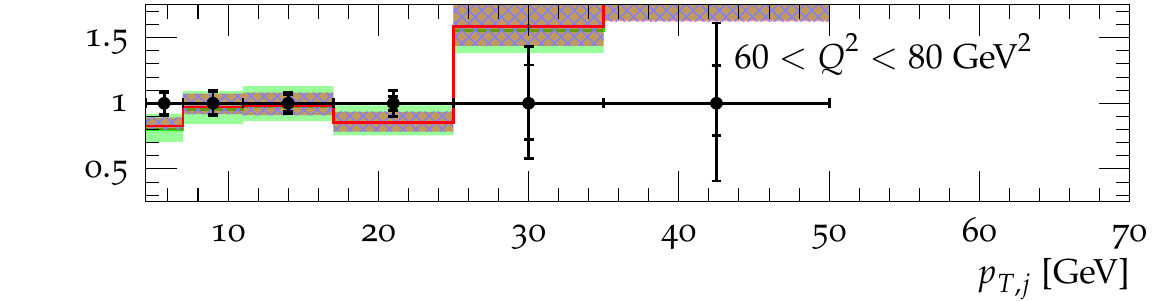}
  \end{minipage}\hfill
  \begin{minipage}{0.325\textwidth}\vskip 4mm
    \includegraphics[width=\textwidth]{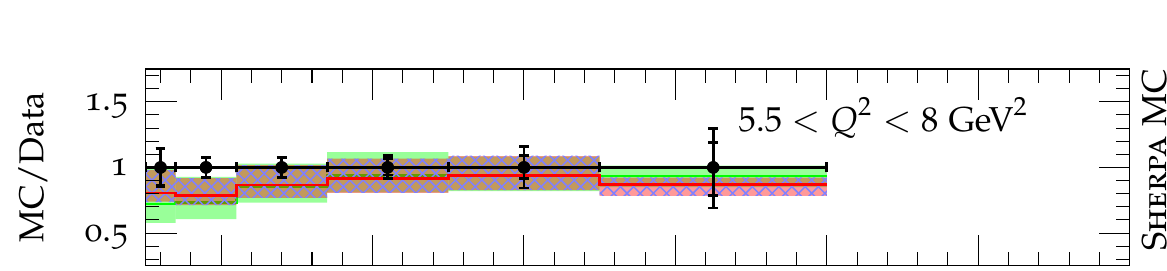}\\[-1mm]
    \includegraphics[width=\textwidth]{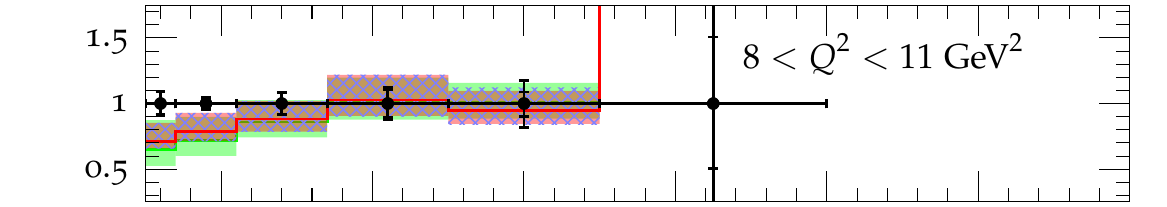}\\[-1mm]
    \includegraphics[width=\textwidth]{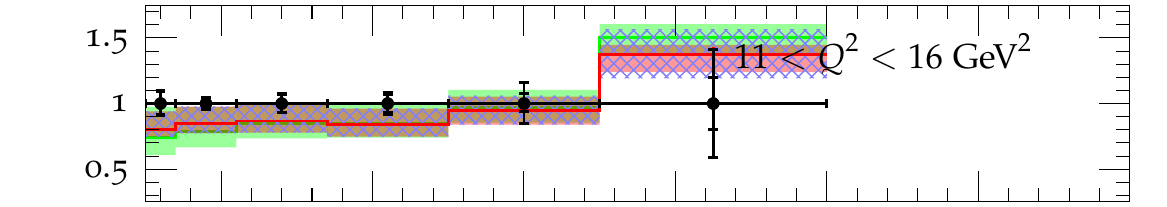}\\[-1mm]
    \includegraphics[width=\textwidth]{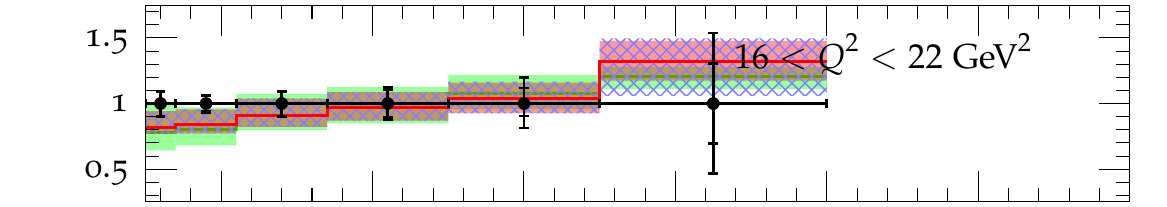}\\[-1mm]
    \includegraphics[width=\textwidth]{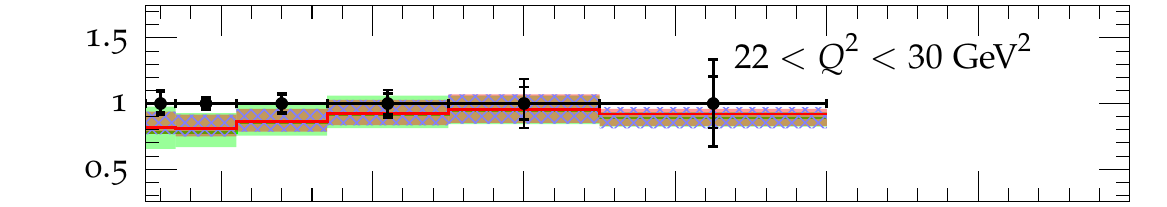}\\[-1mm]
    \includegraphics[width=\textwidth]{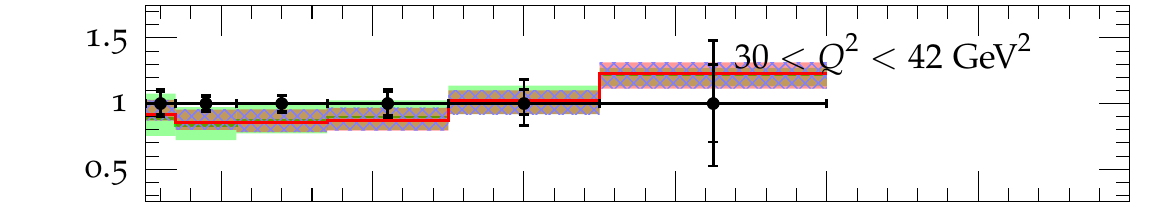}\\[-1mm]
    \includegraphics[width=\textwidth]{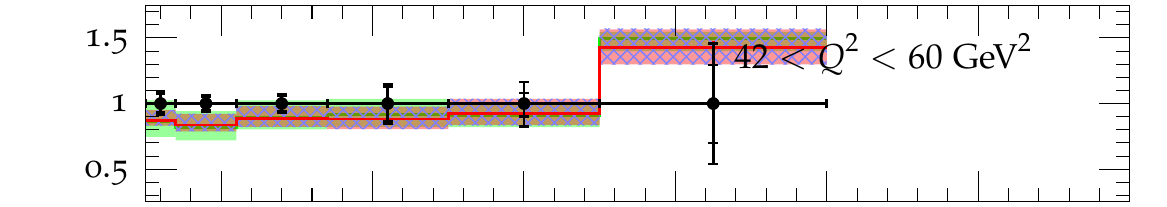}\\[-1mm]
    \includegraphics[width=\textwidth]{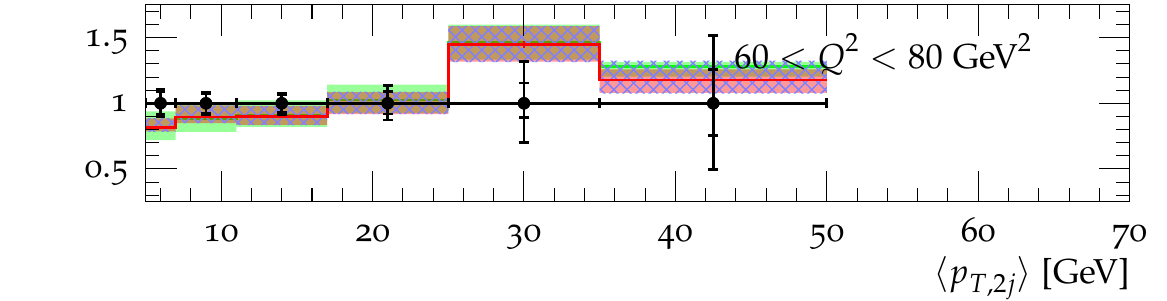}
  \end{minipage}\hfill
  \begin{minipage}{0.325\textwidth}\vskip 4mm
    \includegraphics[width=\textwidth]{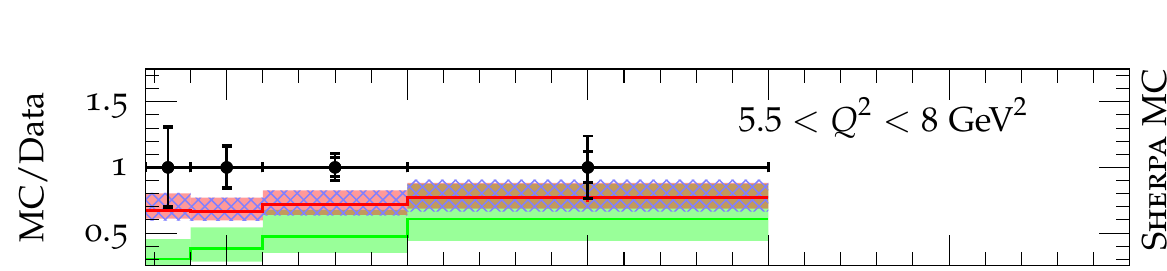}\\[-1mm]
    \includegraphics[width=\textwidth]{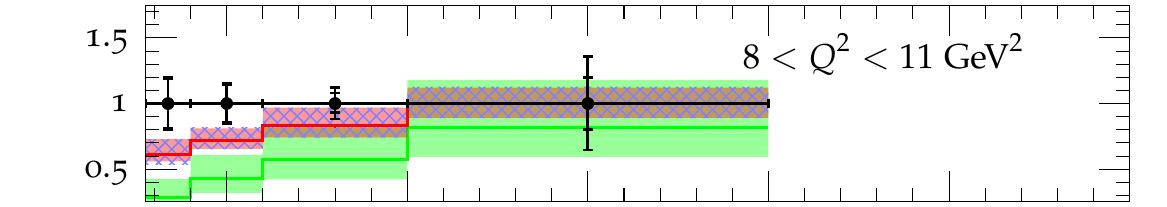}\\[-1mm]
    \includegraphics[width=\textwidth]{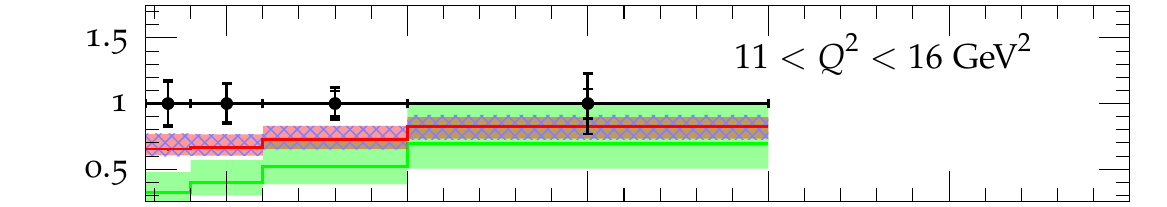}\\[-1mm]
    \includegraphics[width=\textwidth]{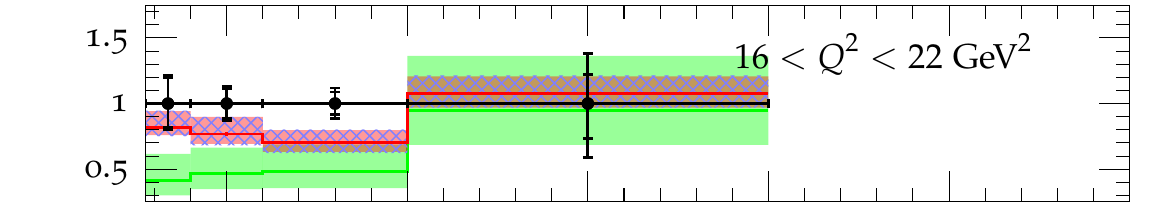}\\[-1mm]
    \includegraphics[width=\textwidth]{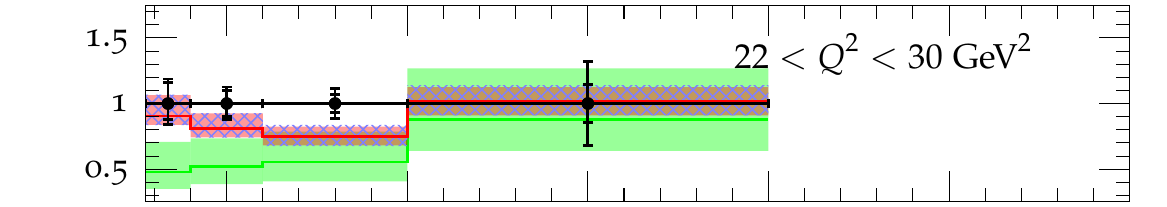}\\[-1mm]
    \includegraphics[width=\textwidth]{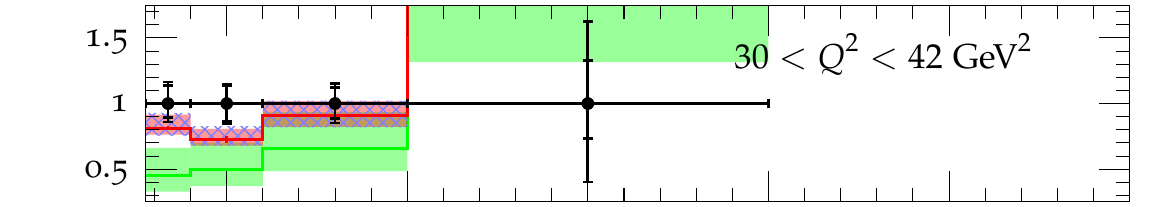}\\[-1mm]
    \includegraphics[width=\textwidth]{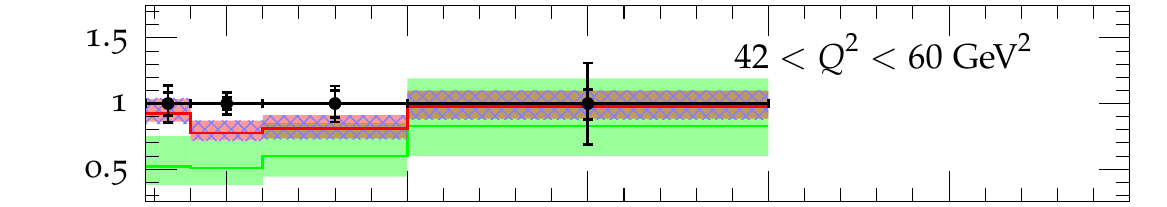}\\[-1mm]
    \includegraphics[width=\textwidth]{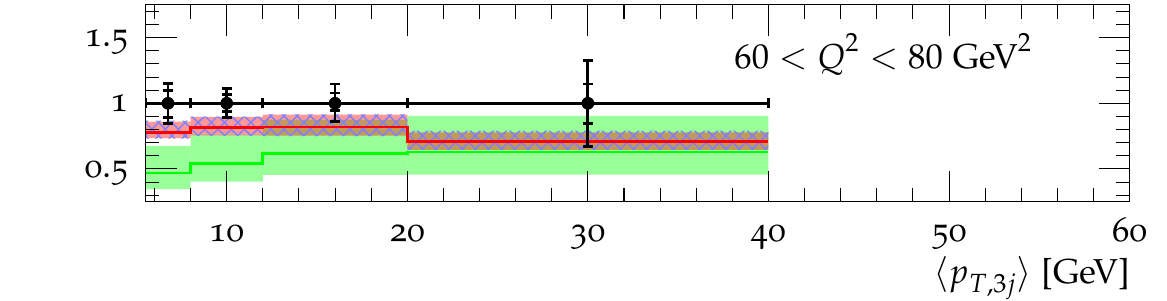}
  \end{minipage}
  \caption{Inclusive jet cross section differential in $Q^2$ as a function
    of $p_{T,j}$ compared to experimental data from the H1 collaboration
    \cite{Andreev:2016tgi}. See Fig.~\ref{fig:inclusive_jetpt_highq2}
    and the main text for details.
    \label{fig:inclusive_jetpt_lowq2}}
\end{figure}

Figures~\ref{fig:inclusive_jetpt_highq2} and~\ref{fig:inclusive_jetpt_lowq2}
show the inclusive jet cross section, the di-jet cross section
and the tri-jet cross section in the high and low $Q^2$ region
as predicted by our simulations in comparison to experimental data
from \cite{Andreev:2014wwa,Andreev:2016tgi}. The shaded uncertainty bands
display the estimated theoretical uncertainties at fixed order and
are obtained from a correlated variation of the renormalization and
factorization scale by factors of two up and down. The hatched uncertainty
band is a combination of the fixed-order uncertainties and the estimated
parton-shower uncertainty, which is obtained by varying the scales at which
the strong coupling and the PDF are evaluated in the parton shower by factors
of $\sqrt{2}$ up and down, using the technique in~\cite{Bendavid:2018nar}.
The parton-shower starting scale $\mu_Q$ is set to $Q^2$. Variations of this scale
have a minor effect on our predictions, because the dominant hierarchy in
the measurement phase space is such that $p_T>Q$.
The predictions shown in green are obtained from a fixed-order NLO
calculation for inclusive jet production.
They have the same fixed-order accuracy as our parton
shower matched simulations because the Born configurations of
inclusive DIS do not contribute to the observables at fixed
order. In the fixed-order predictions, we account for hadronization effects
using the correction factors tabulated
in~\cite{Andreev:2014wwa,Andreev:2016tgi}. They are obtained in the
usual fashion, i.e. by comparing leading-order parton shower Monte Carlo
simulations before and after hadronization. The corresponding ratios
are typically applied as multiplicative corrections to
fixed-order calculations at the level of the observables.
The predictions shown in red are obtained using our \unlops matched calculation
supplemented with the string hadronization model~\cite{Andersson:1983ia,Andersson:1998tv}.
We observe good agreement between the matched calculation and
the fixed-order prediction, which indicates that the QCD evolution
and hadronization effects are well under control. Our results test and
confirm, for the first time at this level of theoretical precision,
the validity of the approach outlined above, where hadronization
corrections are extracted from parton shower Monte Carlo simulations
and then multiplicatively applied to the fixed-order calculation.

\begin{figure}[p]
  \centering
  \begin{minipage}{0.325\textwidth}\vskip 4mm
    \includegraphics[width=\textwidth]{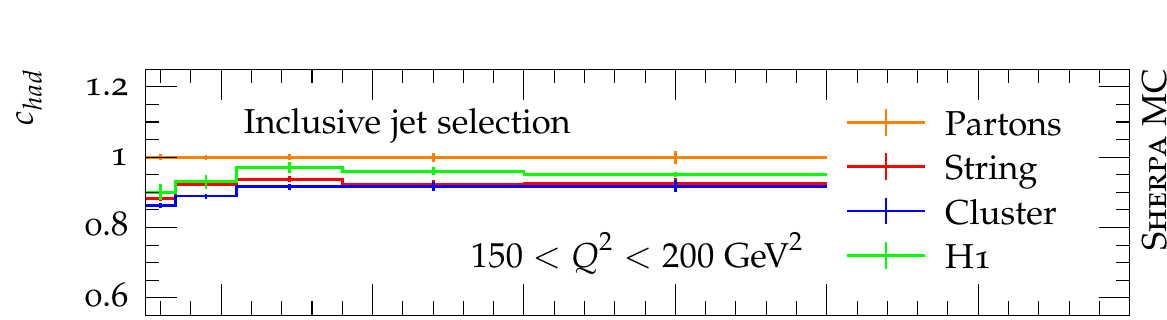}\\[-1mm]
    \includegraphics[width=\textwidth]{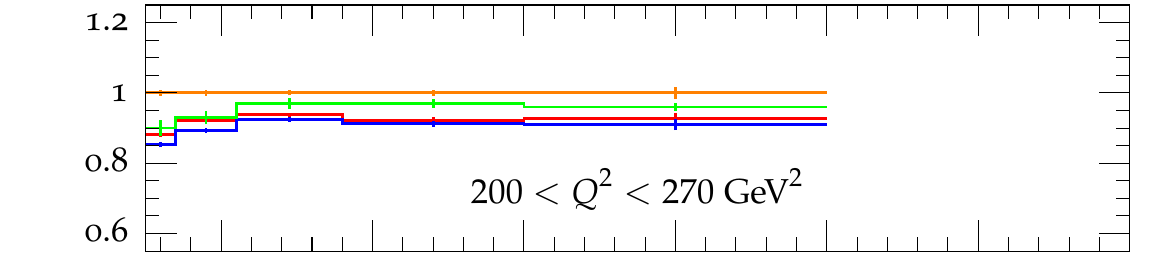}\\[-1mm]
    \includegraphics[width=\textwidth]{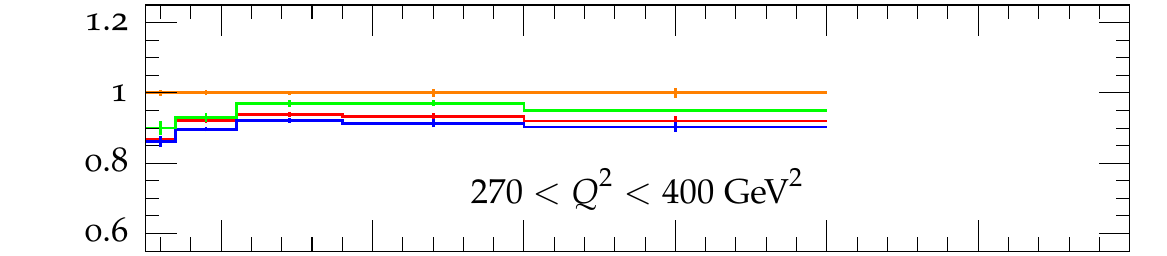}\\[-1mm]
    \includegraphics[width=\textwidth]{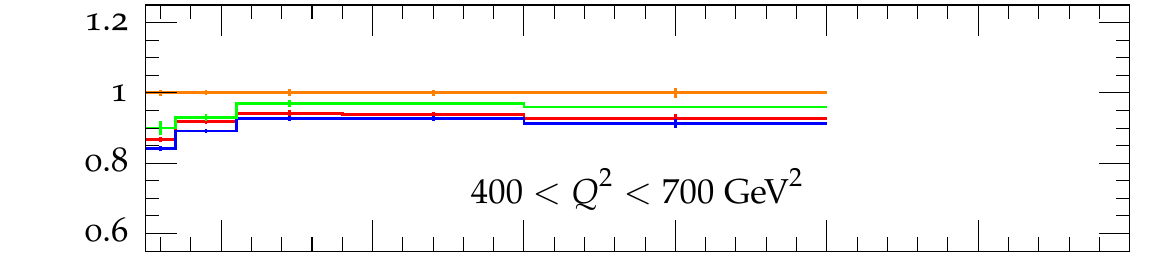}\\[-1mm]
    \includegraphics[width=\textwidth]{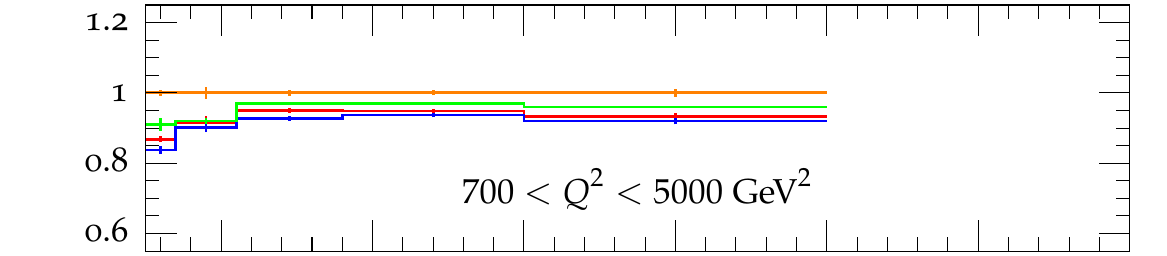}\\[-1mm]
    \includegraphics[width=\textwidth]{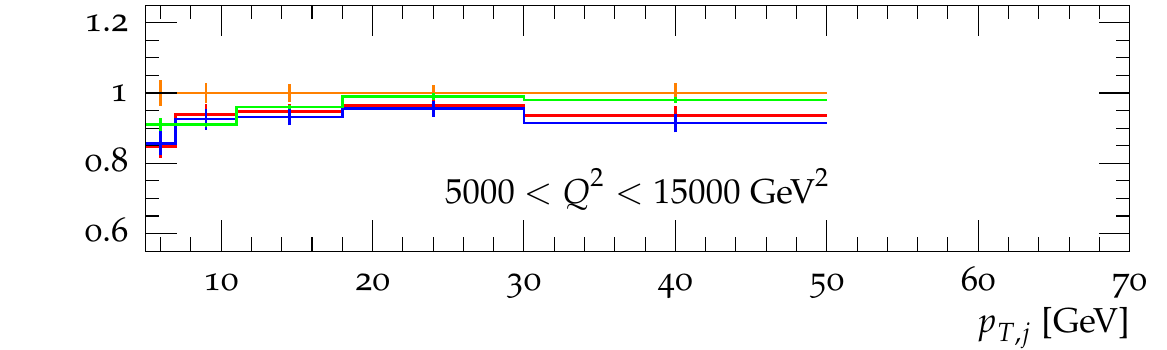}
  \end{minipage}\hfill
  \begin{minipage}{0.325\textwidth}\vskip 4mm
    \includegraphics[width=\textwidth]{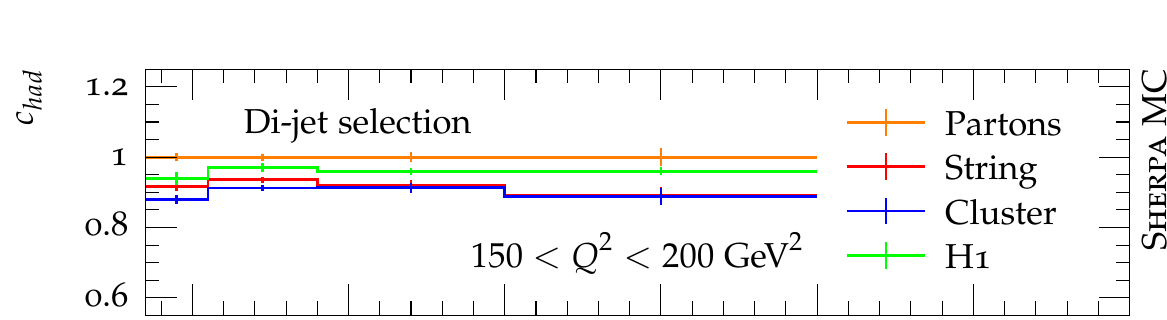}\\[-1mm]
    \includegraphics[width=\textwidth]{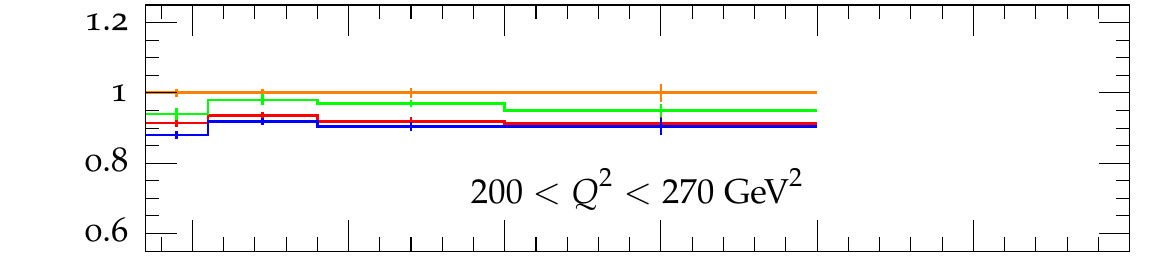}\\[-1mm]
    \includegraphics[width=\textwidth]{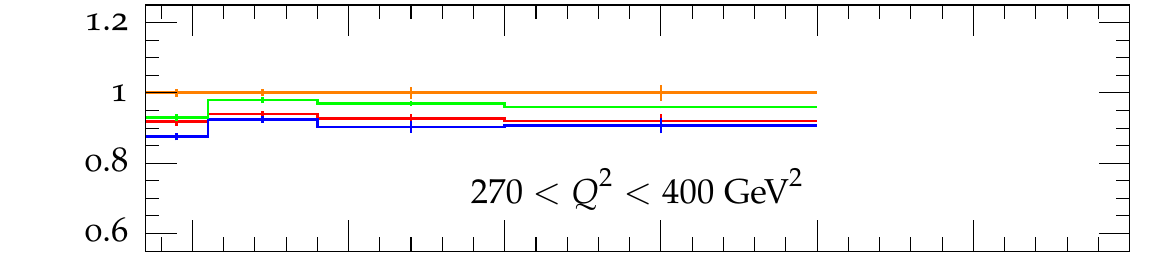}\\[-1mm]
    \includegraphics[width=\textwidth]{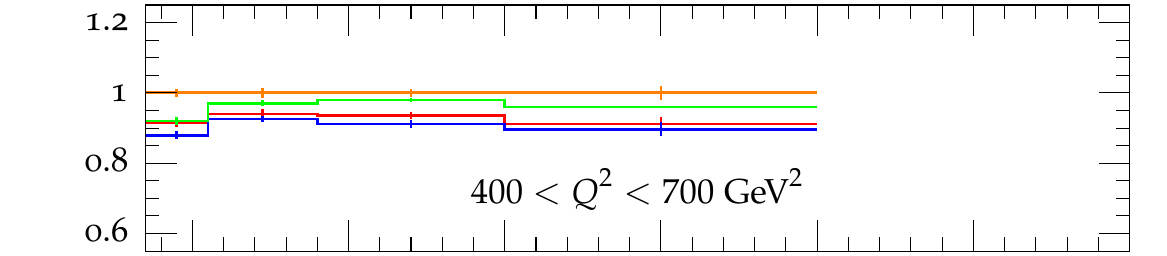}\\[-1mm]
    \includegraphics[width=\textwidth]{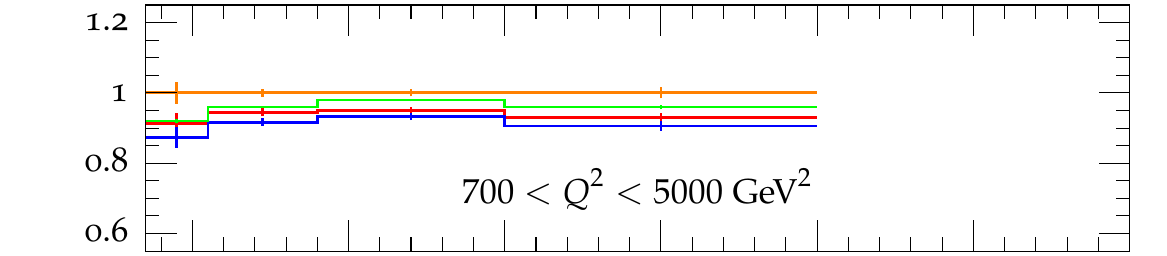}\\[-1mm]
    \includegraphics[width=\textwidth]{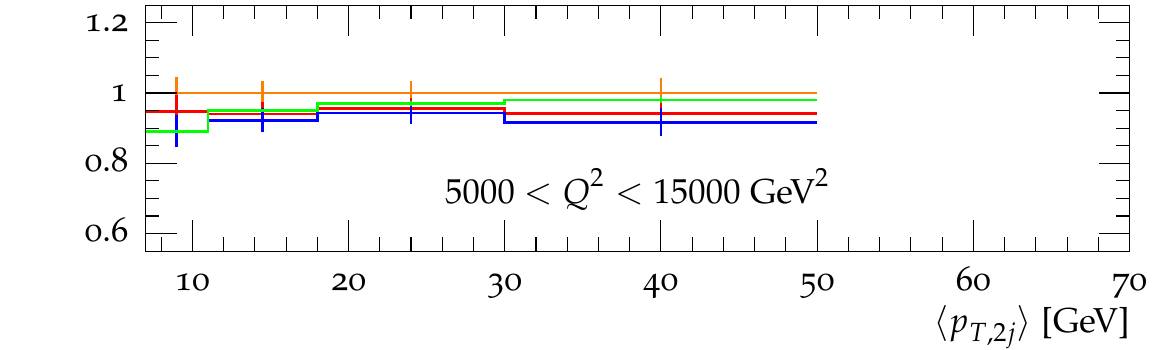}
  \end{minipage}\hfill
  \begin{minipage}{0.325\textwidth}\vskip 4mm
    \includegraphics[width=\textwidth]{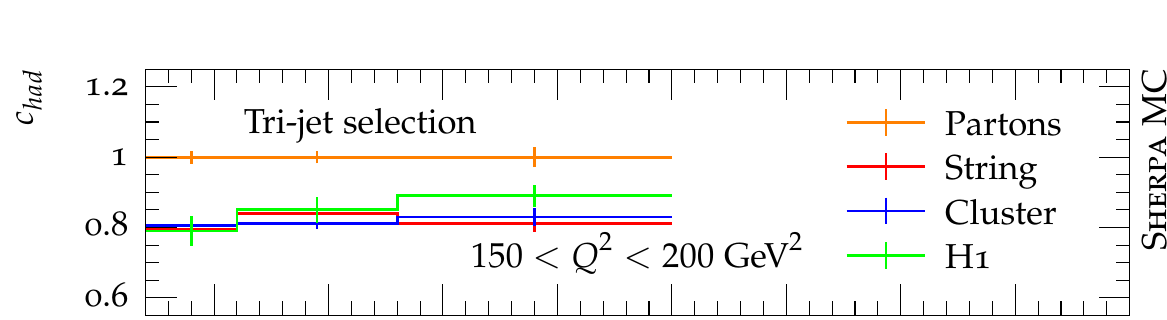}\\[-1mm]
    \includegraphics[width=\textwidth]{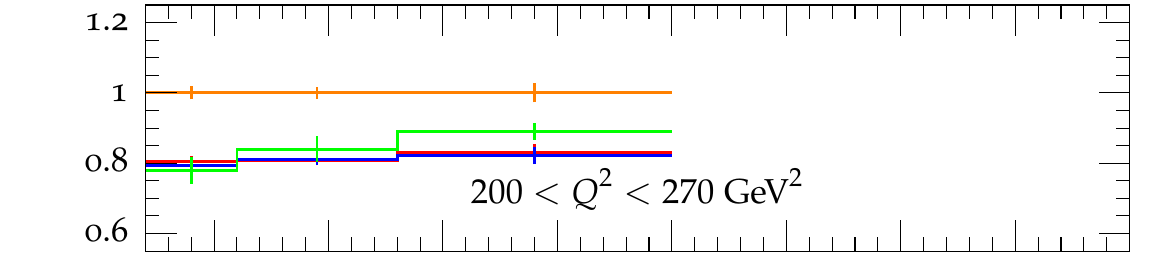}\\[-1mm]
    \includegraphics[width=\textwidth]{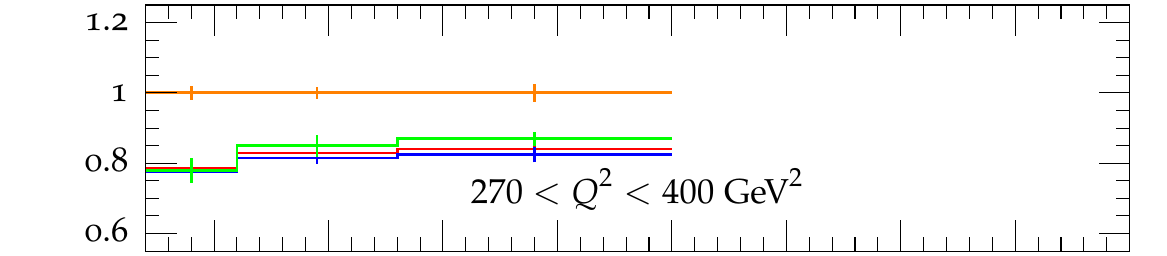}\\[-1mm]
    \includegraphics[width=\textwidth]{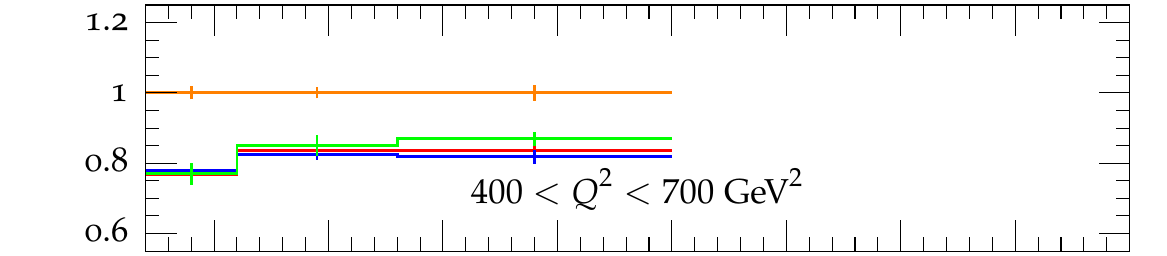}\\[-1mm]
    \includegraphics[width=\textwidth]{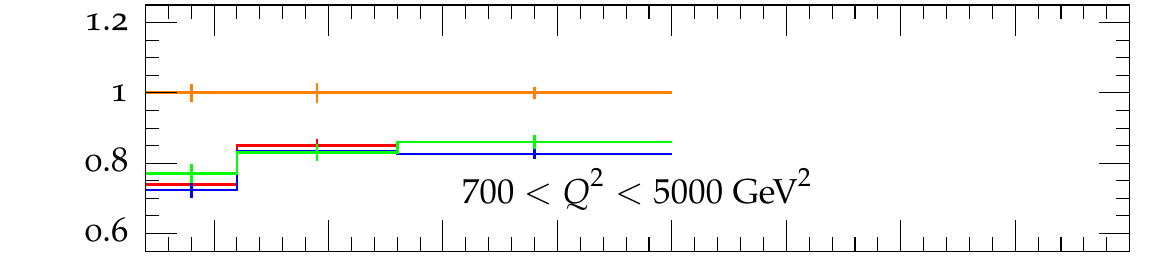}\\[-1mm]
    \includegraphics[width=\textwidth]{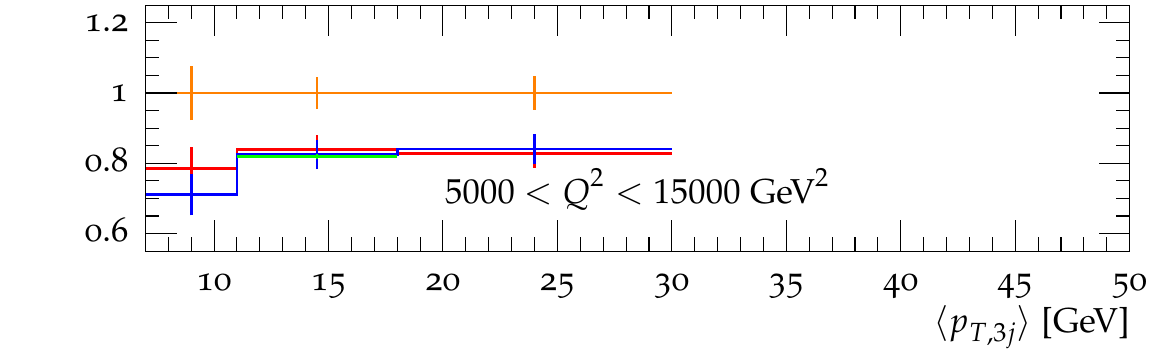}
  \end{minipage}\\
  \begin{minipage}{0.325\textwidth}\vskip 4mm
    \includegraphics[width=\textwidth]{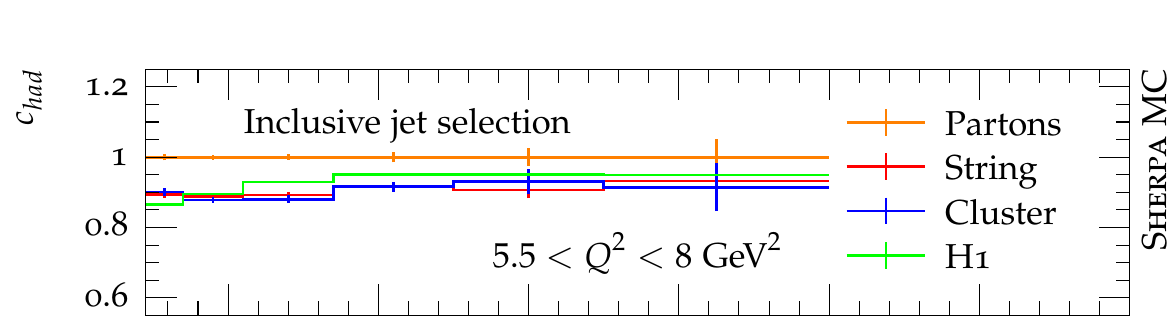}\\[-1mm]
    \includegraphics[width=\textwidth]{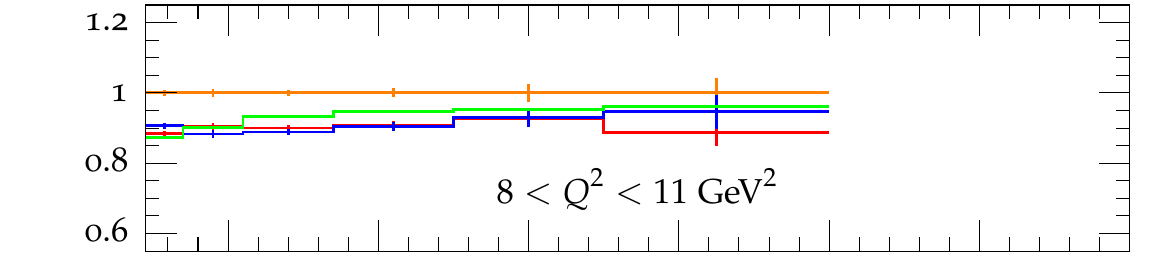}\\[-1mm]
    \includegraphics[width=\textwidth]{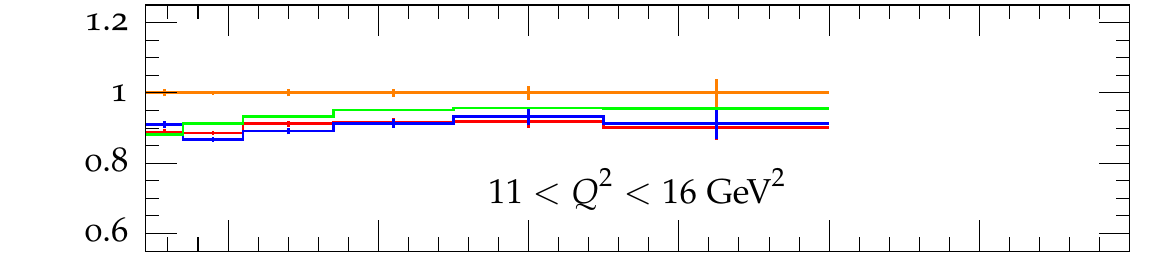}\\[-1mm]
    \includegraphics[width=\textwidth]{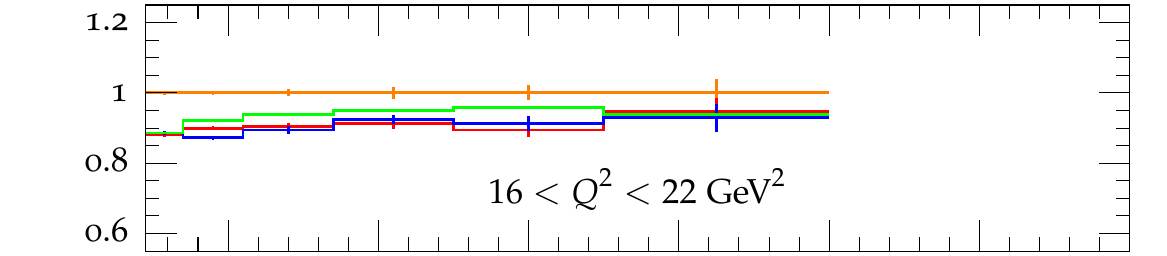}\\[-1mm]
    \includegraphics[width=\textwidth]{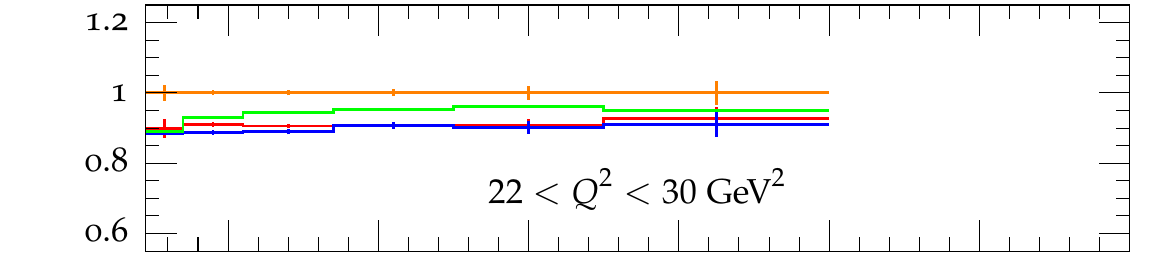}\\[-1mm]
    \includegraphics[width=\textwidth]{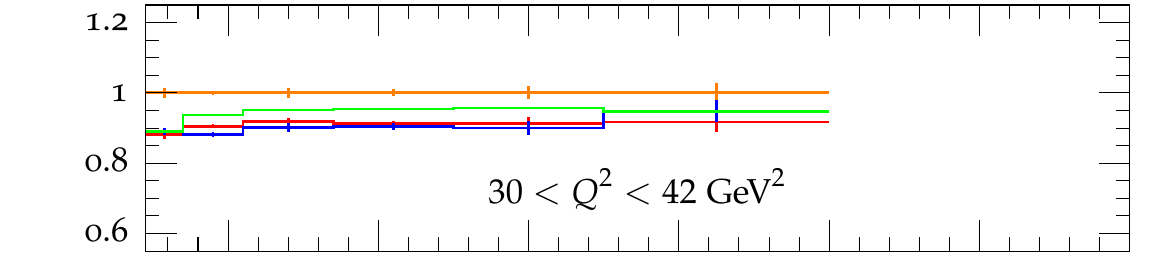}\\[-1mm]
    \includegraphics[width=\textwidth]{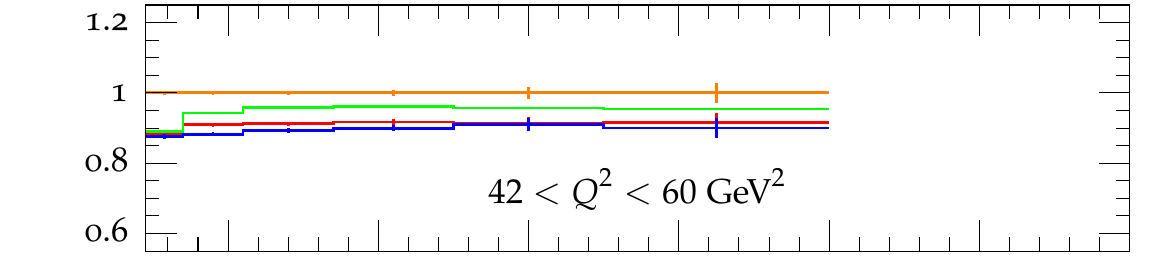}\\[-1mm]
    \includegraphics[width=\textwidth]{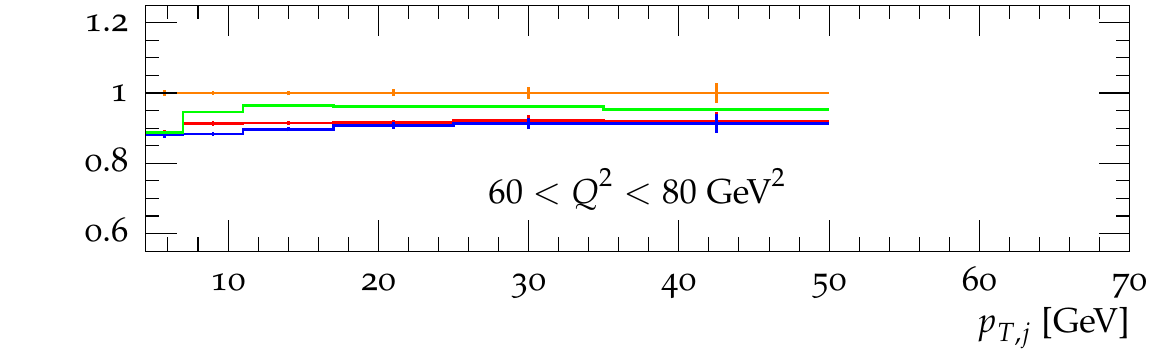}
  \end{minipage}\hfill
  \begin{minipage}{0.325\textwidth}\vskip 4mm
    \includegraphics[width=\textwidth]{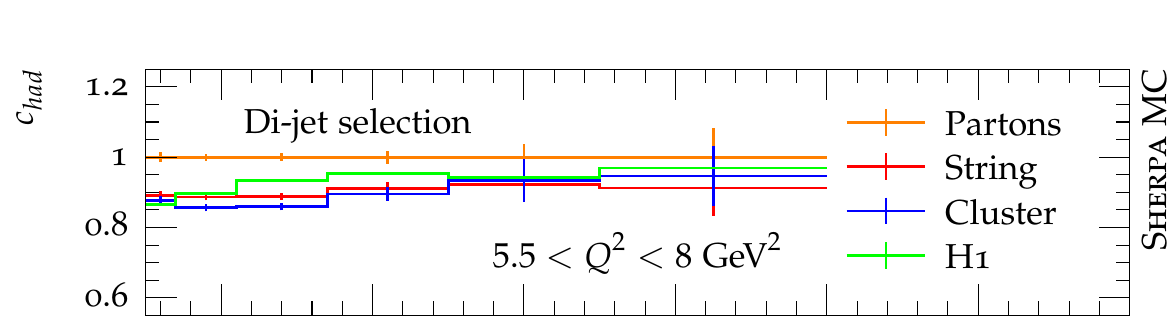}\\[-1mm]
    \includegraphics[width=\textwidth]{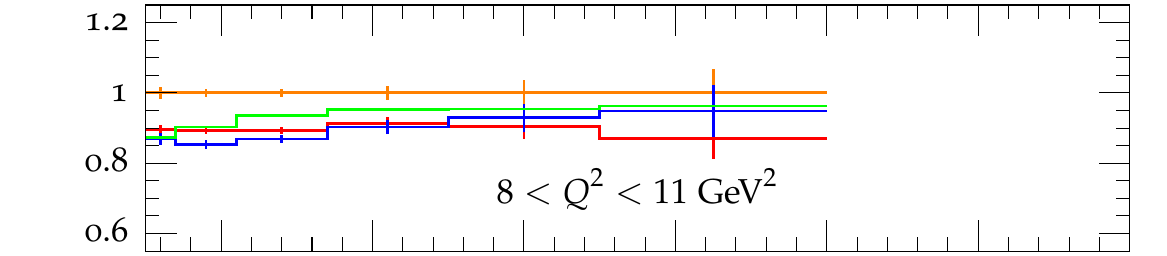}\\[-1mm]
    \includegraphics[width=\textwidth]{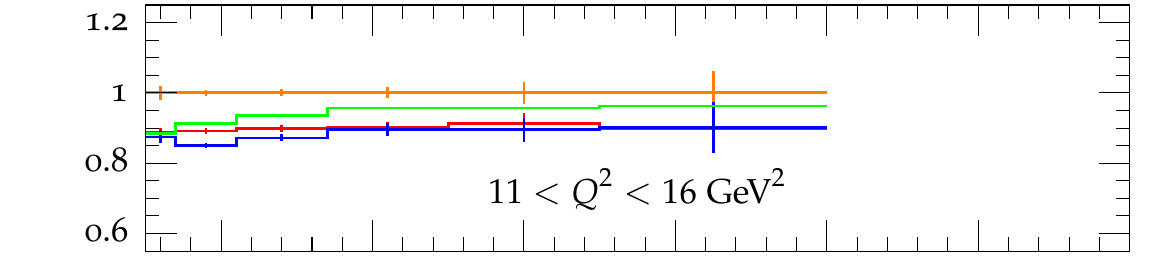}\\[-1mm]
    \includegraphics[width=\textwidth]{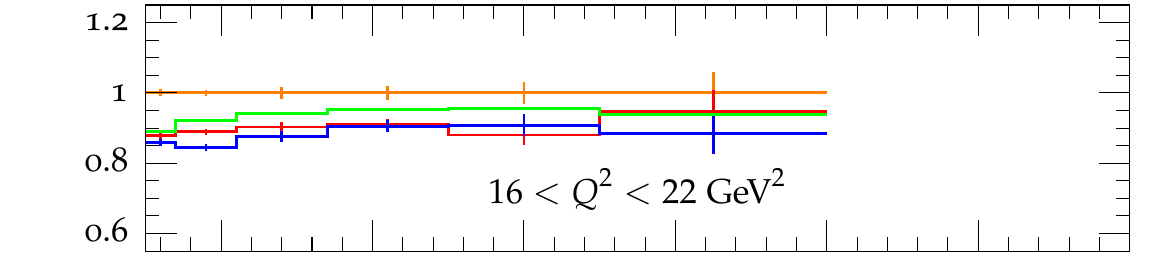}\\[-1mm]
    \includegraphics[width=\textwidth]{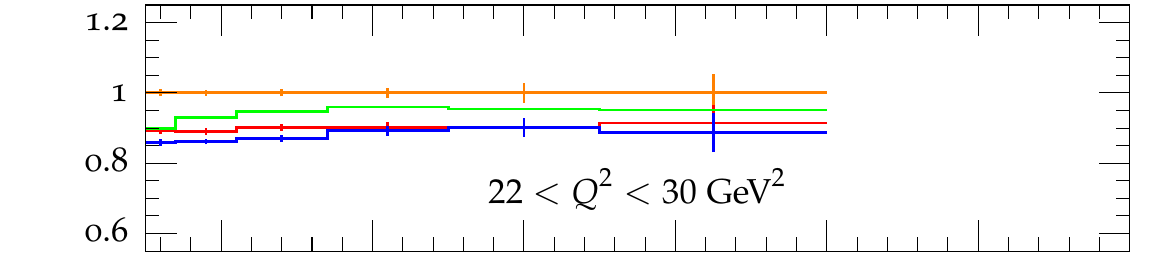}\\[-1mm]
    \includegraphics[width=\textwidth]{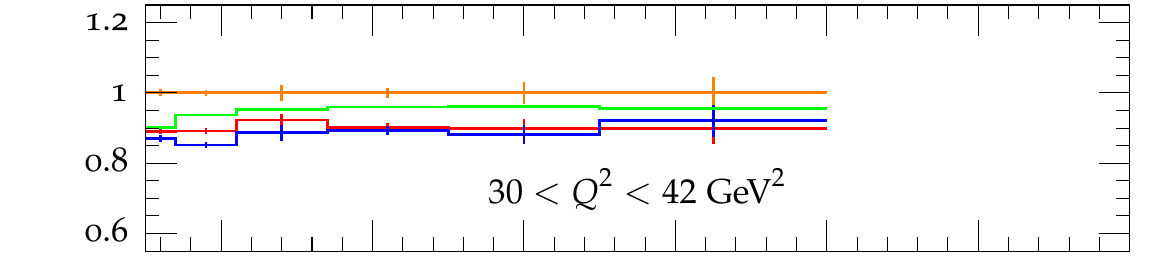}\\[-1mm]
    \includegraphics[width=\textwidth]{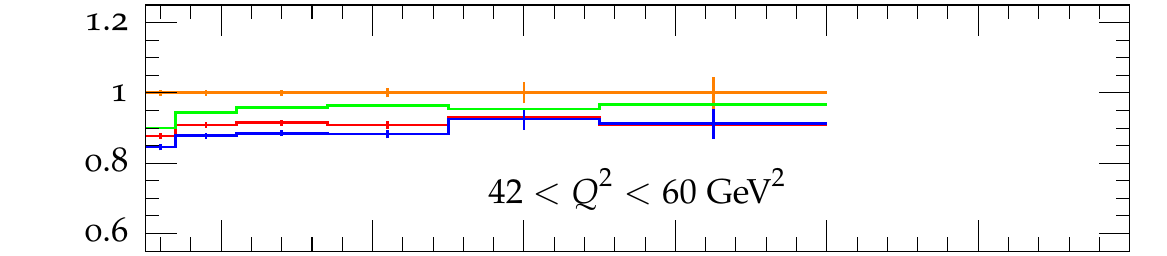}\\[-1mm]
    \includegraphics[width=\textwidth]{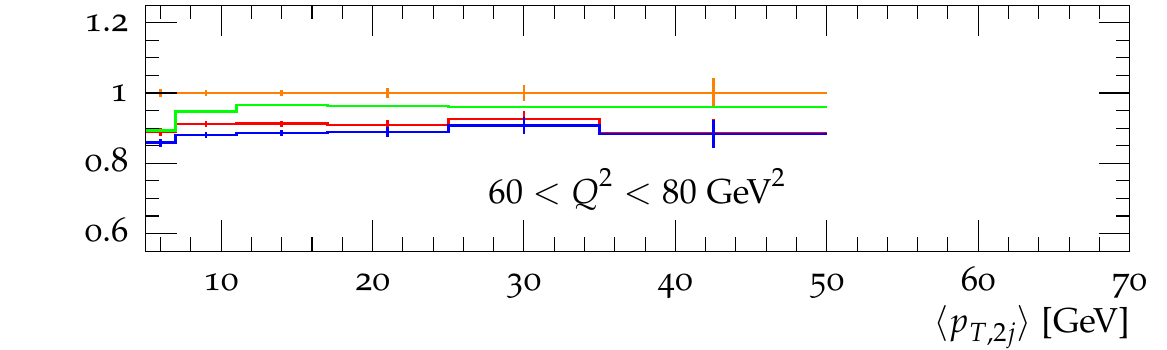}
  \end{minipage}\hfill
  \begin{minipage}{0.325\textwidth}\vskip 4mm
    \includegraphics[width=\textwidth]{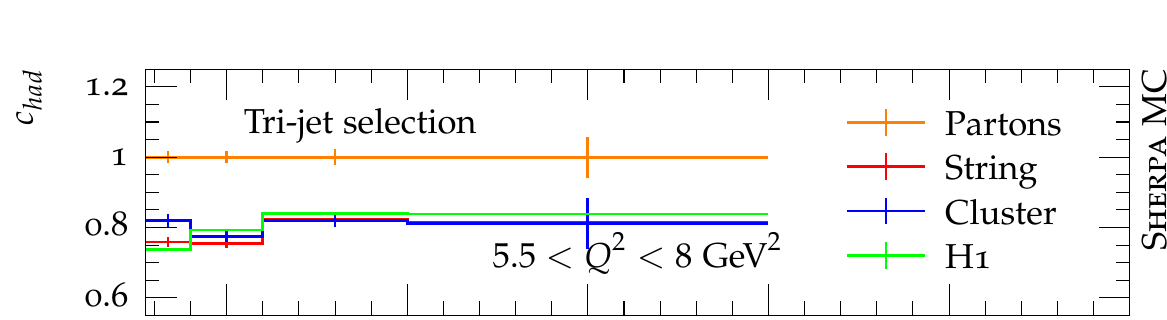}\\[-1mm]
    \includegraphics[width=\textwidth]{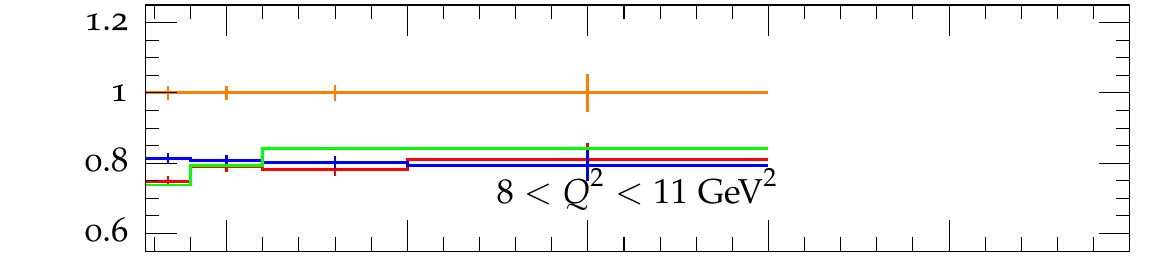}\\[-1mm]
    \includegraphics[width=\textwidth]{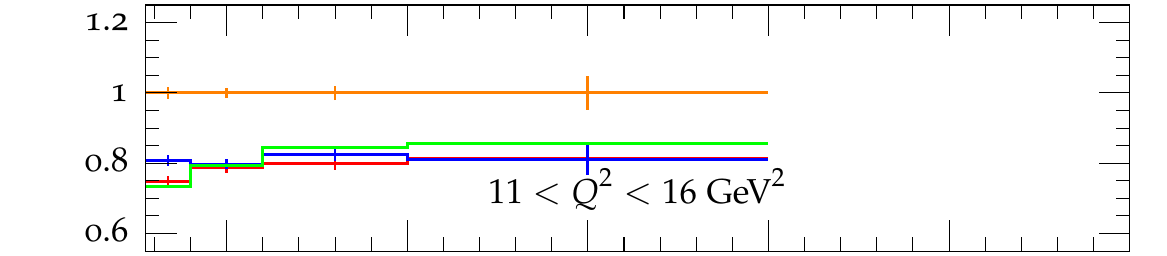}\\[-1mm]
    \includegraphics[width=\textwidth]{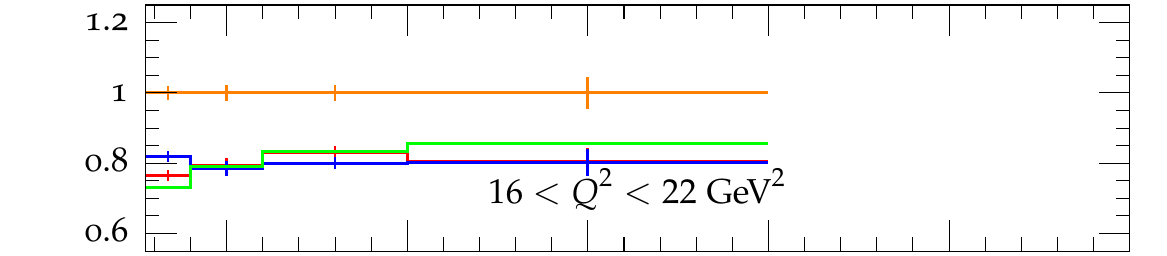}\\[-1mm]
    \includegraphics[width=\textwidth]{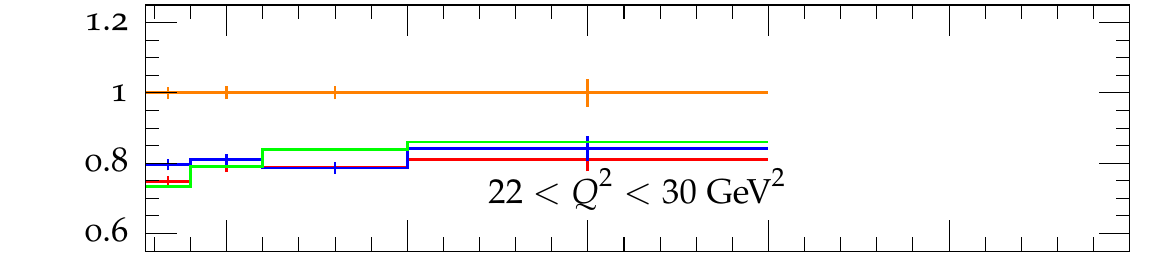}\\[-1mm]
    \includegraphics[width=\textwidth]{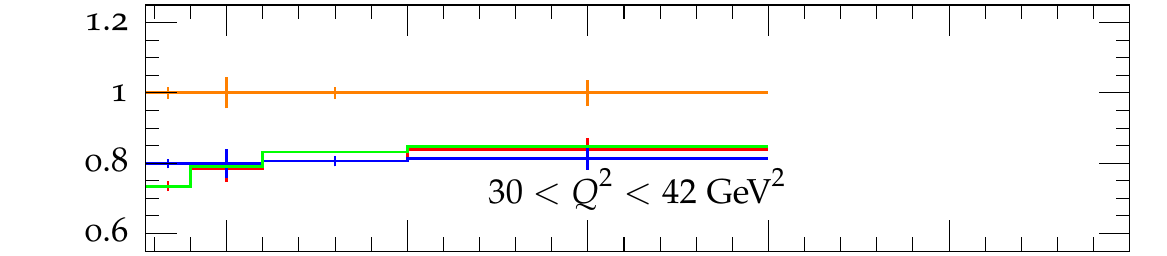}\\[-1mm]
    \includegraphics[width=\textwidth]{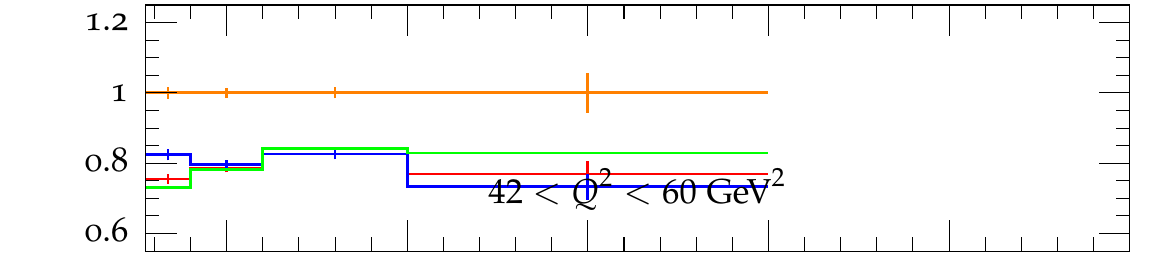}\\[-1mm]
    \includegraphics[width=\textwidth]{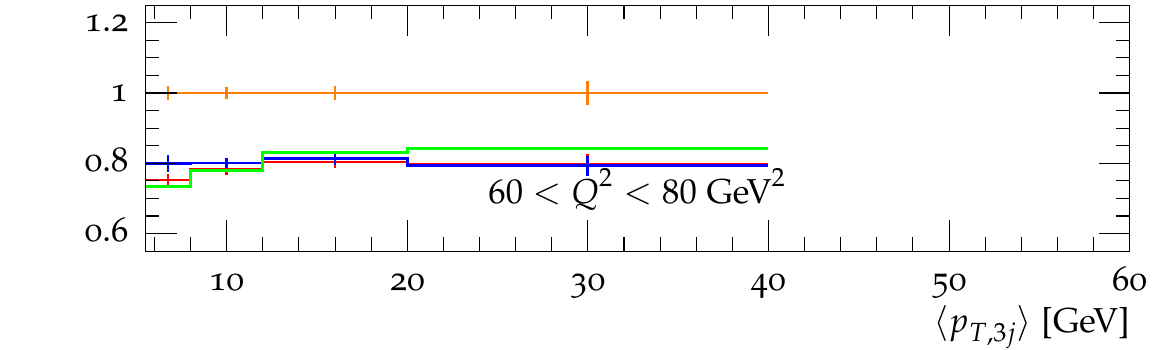}
  \end{minipage}
  \caption{Hadronization corrections determined using the \unlops simulation
    of DIS. Predictions from the Lund string model (red) are compared to
    results from the cluster fragmentation as implemented in \Sherpa (blue)
    and to results computed by the H1 collaboration
    \cite{Andreev:2014wwa,Andreev:2016tgi}.    
    \label{fig:hadronization_corrections}}
\end{figure}

We quantify the size and uncertainty of the hadronization corrections
in our simulations in Fig.~\ref{fig:hadronization_corrections}. We plot
the ratio between the particle-level predictions, with hadronization
and hadron decays applied, to the prediction obtained after parton showering.
We show two different results, one obtained using the Lund string fragmentation
model~\cite{Andersson:1983ia,Andersson:1998tv} as implemented
in \Pythia~6.4~\cite{Sjostrand:2006za} and one obtained using the
cluster fragmentation model~\cite{Field:1982dg,Webber:1983if}
as implemented in \Sherpa~\cite{Winter:2003tt}.
The perturbative input to the two simulations is identical, hence
we quote their difference as the estimated hadronization uncertainty.
We note that the hadronization corrections in our approach agree well
with the results computed by H1, which are based on computations using
\Djangoh~\cite{Charchula:1994kf} and \RapGap~\cite{Jung:1993gf}.
These two generators both include the exact expressions for the QCD Compton process
at leading order, but \Djangoh uses the Lund dipole cascade model of
\Ariadne~\cite{Lonnblad:1992tz} to simulate higher-order radiative corrections,
while \RapGap is based on collinear parton evolution. Both make use
of the Lund string fragmentation model. It is encouraging that in our
approach, using both a higher-order perturbative input and a different
parton-shower model, the size of the hadronization corrections is very similar.
In addition, the hadronization uncertainties are small, except for the
very low transverse momentum region in the tri-jet cross section.

\section{Conclusions}
\label{sec:conclusions}
We have presented the first parton-shower matched calculation
of hadronic final state production in DIS at NNLO QCD precision.
The techniques needed to perform the simulation are implemented
in the publicly available program \Sherpa and can be used
for event generation at the particle level.
In contrast to earlier calculations using non-unitary
multi-jet merging techniques, we are able to predict both jet
production rates and inclusive quantities like structure functions
within a single calculation. The agreement with recent analyses
by the H1 collaboration is good. Since our calculation is based
on collinear factorization, it can provide the basis for the
reliable extraction of hadronization corrections needed in the
comparison to fixed-order calculations.

\begin{acknowledgments}
  We thank Daniel Britzger and Thomas Gehrmann for many helpful discussions
  and for their comments on the manuscript. We would also like to thank the authors
  of the Apfel library and HERAFitter (in particular Valerio Bertone,
  Voica Radescu and Ringaile Placakyte) for their support. This
  work was supported by the US Department of Energy under contract
  DE--AC02--76SF00515.
\end{acknowledgments}

\clearpage
\bibliography{journal}
\end{document}